\documentclass[preprint]{elsarticle}

\usepackage[utf8]{inputenc}
\usepackage{graphicx}
\usepackage{amsmath,amssymb}
\usepackage{rotating}
\usepackage{dsfont}

\usepackage[margin=2cm]{geometry}

\newcommand{\HHHOO}{{H$_3$O$_2$}$^-$}
\newcommand{\eg}{e.g.\ }
\newcommand{\ie}{i.e.\ }
\newcommand{\cf}{cf.\ }
\newcommand{\kB}{k_\text{B}}
\newcommand{\ic}{\text{cm}^{-1}}
\newcommand{\dg}{\ensuremath{^\circ}}
\newcommand{\vct}[1]{\boldsymbol{#1}}
\newcommand{\refeq}[1]{Eq.\ (\ref{#1})}

\journal{Chemical Physics}

\begin{document}

\begin{frontmatter}

\title{
Accuracy of Potfit-based potential representations and its
impact on the performance of (ML-)MCTDH}

\author{Frank Otto}
\ead{oft.kontra@gmail.com}
\address{Dept. of Chemistry, University College London,
        20 Gordon St., London WC1H 0AJ, UK}

\author{Ying-Chih Chiang}
\address{Dept. of Physics, The Chinese University of Hong Kong, Sha Tin, N.T., Hong Kong}

\author{Daniel Peláez}
\address{Laboratoire de Physique des Lasers, Atomes et Molécules (PhLAM), Unité Mixte de Recherche (UMR) 8523,
         Université Lille 1, Bât. P5, Villeneuve d’Ascq Cedex, France}

\begin{abstract}
Quantum molecular dynamics simulations with MCTDH or ML-MCTDH perform
best if the potential energy surface (PES) has a sum-of-products (SOP) or
multi-layer operator (MLOp) structure. Here we investigate four different
POTFIT-based methods for representing a general PES as such a structure,
among them the novel random-sampling multi-layer Potfit (RS-MLPF). We
study how the format and accuracy of the PES representation influences
the runtime of a benchmark (ML-)MCTDH calculation, namely the computation
of the ground state of the \HHHOO{} ion. Our results show that compared to
the SOP format, the MLOp format leads to a much more favorable scaling
of the (ML-)MCTDH runtime with the PES accuracy. At reasonably high
PES accuracy, ML-MCTDH calculations thus become up to 20 times faster,
and taken to the extreme, the RS-MLPF method yields extremely accurate
PES representations (global root-mean-square error of $\sim 0.1\,\ic$) which
still lead to only moderate computational demands for ML-MCTDH.
\end{abstract}


\end{frontmatter}

\section{Introduction}

The computational treatment of high-dimensional quantum systems
unavoidably necessitates the use of approximations, given the currently
available hardware facilities.  These approximations include the
discretization of the Hilbert space (using grid-based or basis-set methods),
the representation of the wavefunction, and the time-stepping method
(for dynamical problems).  In the case of quantum molecular dynamics (QMD),
which is usually carried out in the framework of the Born-Oppenheimer
or (in the case of several electronic states) group Born-Oppenheimer approximations \cite{wor04:127},
further approximations are introduced by the creation of one or more
potential energy surfaces (PES) on which the nuclei move, which involves
the level of electronic structure theory, the electronic basis set, and how the computed
electronic energies are fitted to a global function.
Moreover, to make use of such a PES in the QMD calculations, further approximating transformations of
the PES may be needed.
Alternatively, promising on-the-fly QMD methods are available as well
\cite{worth_novel_2004, worth_solving_2008}, though the computational effort
in the electronic structure calculations still limits their general applicability.
In all these approximation steps, the balance between accuracy and computational
resource usage must be considered, and it is worth noting that different
approximation methods exhibit different scalings between approximation
accuracy and resource usage.
Since researchers are usually faced with limited computational budgets, 
ideally the approximation methods should be chosen such that resource
usage is minimized for a given desired accuracy.

Concerning the representation of the wavefunction, major breakthroughs in
obtaining accurate results with limited resources were achieved by
Meyer and coworkers in developing
the multi-configuration time-dependent Hartree method (MCTDH)
\cite{mey90:73,bec00:1,mey09:book,mey12:351}.  MCTDH has proven
to be a powerful computational method for high-dimensional quantum systems,
both for investigating the system dynamics
\cite{raa99:936,cat01:2088,gat05:174311,ven07:184303,ven08:4692,ott08:064305,ott12:619}
as well as for determining spectroscopic features
\cite{mar05:204310,ven07:184302,ven07:184303,ven09:234305,sch11:234307,sch14:034116,sad14:114101,pel14:42,pel17:100},
where it should be noted that the list of cited works is far from complete.
With its multi-layer extension ML-MCTDH
\cite{wan03:1289,man08:164116,ven11:044135},
accurate full-dimensional quantum calculations for systems with up to hundreds
of degrees of freedom (DOFs) are possible
\cite{wan03:1289,cra07:144503,wan10:78,wes11:184102,wel12:244106,cra11:064504,ven11:044135,men12:134302,men13:014313,xie15:084706}.
From the approximation perspective, (ML-)MCTDH is particularly
attractive as it allows to systematically increase the accuracy by increasing
the size of the wavefunction representation, at the expense of using more
computational resources. In this respect ML-MCTDH fares better than MCTDH,
as resource usage depends only polynomially instead of exponentially
on the wavefunction size \cite{man08:164116,ott14:014106}.

A crucial point for the efficient application of (ML-)MCTDH is that the PES
needs to be represented in a
suitable format which facilitates the efficient evaluation of the various high-dimensional
integrals that are encountered in the course of the (ML-)MCTDH algorithm.
This requirement can be overcome by using the CDVR method \cite{man96:6989,man09:054109}
which however requires a very large amount of PES evaluations,
and considerable development effort may be needed to overcome
the resulting bottleneck by massively speeding up the PES routine \cite{wel13:164118}.
In general it is much more economical to represent the PES as a sum of products of one- or
low-dimensional potential functions, an approach that has been traditionally used with MCTDH,
as it transforms each high-dimensional integral into a sum of products of one- or low-dimensional
integrals, which are easily evaluated.
This sum-of-products (SOP) format is a suitable choice for ML-MCTDH as well \cite{man08:164116}.
More recently, one of the authors has shown that for ML-MCTDH there exists an
alternative representation of the PES as a \emph{hierarchical} sum of products, i.e.
as a \emph{multi-layer operator} (MLOp), which allows for a much more compact
representation of the PES and hence for a much more efficient evaluation of the
ML-MCTDH equations of motion \cite{ott14:014106}. 

For most model systems, the PES can be naturally expressed in SOP format.
However, general PES are usually given as a high-dimensional analytic function
fitted to a large number of electronic structure calculations,
and for such PES it is necessary to transform them into SOP or MLOp format.
In general, it is not possible to perform this transformation exactly, but only
approximately; and more accurate representations of the PES unavoidably
lead to a larger number of terms in the representation, which negatively impacts
the performance of (ML-)MCTDH (\eg for the SOP format, the computational effort
for MCTDH is directly proportional to the number of summands in the expansion).
Given a desired accuracy of the PES representation, one should then strive to
find a PES representation which minimizes the effort for (ML-)MCTDH.

There are a number of options for performing this transformation, most of them
targeting the SOP format.  
One such approach makes
use of neural networks with exponential activation functions
\cite{man06:084109, pradhan_ab_2013, manzhos_neural_2015, pradhan_ground_2017, brown_fitting_2017}
but so far this method has only been demonstrated for
fitting functions up to 6D (\ie up to four atoms).
Another approach is the $n$-mode representation ($n$-MR) \cite{car97:10458}, also known as
cut high-dimensional model representation (cut-HDMR) \cite{rab99:197},
which expands the PES as a sum of one- and few-body terms, and the latter
can be transformed into SOP format relatively easily.
This method can scale to rather large systems, and it has been used
in MCTDH calculations \eg on the Zundel cation
\cite{ven07:6918,ven07:184302,ven07:184303,ven09:234305,ven09:034308} 
as well as on malonaldehyde \cite{sch11:234307},
and it has also been used together with the neural networks approach \cite{manzhos_using_2008}.
However, the drawback of the $n$-MR approach is that the number of terms
grows exponentially with the expansion order, so that for larger systems
one must choose (usually using statistical methods) the most relevant
terms in the $n$-MR expansion. This makes the method non-variational, \ie
adding more terms to the expansion does not necessarily increase its accuracy,
so that it is difficult to obtain a PES fit with a prescribed accuracy.
Furthermore, it has been shown that this non-variational behaviour
may lead to artificial structure in the PES \cite{kro07:334}.
In contrast, the Potfit method \cite{jae96:7974} offers good control
over the accuracy of the SOP fit, as it is variational. Given the values of the
PES on a product grid, Potfit determines optimal one-dimensional potential
functions for each DOF, and expresses the PES as a linear combination of
products of these potential functions (technically, a truncated Tucker
decomposition \cite{tuc63:122}), which results in a SOP expansion close to optimal.
However, the requirement of knowing
the PES on a full product grid limits the applicability of Potfit to systems
with 6--8 DOFs.  This limitation can be overcome by approximating the
integrations over the full grid by multi-grid or Monte Carlo methods, giving
rise to the multi-grid Potfit (MGPF) \cite{pel13:014108} and the Monte Carlo Potfit
(MCPF) \cite{mcpf} methods, respectively.
MGPF is expected to be able to treat PES up to 12D, while
MCPF has already been applied to a 15D PES, though with considerable
computational effort.
On the other hand,
for transforming a PES into MLOp format there currently
exists to our knowledge only one method, multi-layer Potfit (MLPF) \cite{ott14:014106},
which is based on the hierarchical singular value decomposition \cite{grasedyck_hierarchical_2010}.
As initially conceived, MLPF like Potfit requires knowing the PES on a full product grid,
limiting its applicability.  To overcome this problem,  
the present article introduces a novel method for transforming a PES into MLOp form,
which avoids having to evaluate the PES on the full product grid by using ideas
originally introduced in MCPF \cite{mcpf}.
The resulting method is termed ``random sampling multi-layer Potfit'' (RS-MLPF),
and by design we expect it to scale to systems as large as those
treatable by MCPF and likely beyond.

In the present work, we aim to investigate how the accuracy of the 
PES fit into SOP/MLOp format influences the runtime of 
(ML-)MCTDH calculations carried out with this fit.
For this purpose we have chosen the \HHHOO{} ion as a benchmark
system, for which we compute the ground state and its zero-point energy
using relaxation methods.
This 9-dimensional (9D) system is slightly too large to be treated with
the original Potfit method, but PES fits suitable for (ML-)MCTDH can be
obtained with the aforementioned improved Potfit variants.
In particular, we use four such variants in our calculations:
(i) MGPF,
(ii) MGPF postprocessed by MLPF,
(iii) MLPF directly, and
(iv) the newly developed RS-MLPF.
For each variant, we produce a series of PES fits
at different accuracy levels, and measure the resulting runtime for
the (ML-)MCTDH relaxation.  The accuracy of each fit is assessed by
sampling the fitting error with Monte Carlo as well as classical molecular
dynamics methods.
Our results reveal that the choice of
Potfit variant can have a strong influence on the computational
resource usage of (ML-)MCTDH, especially when a highly accurate
PES fit is required.

The article is structured as follows:
Section \ref{sec:methods} reviews the methods used in this study.
In particular, Section \ref{sec:rs-mlpf} gives details on the novel RS-MLPF method.
Section \ref{sec:h3o2} describes some properties of the \HHHOO{} ion and the PES used.
Our calculations and their results are given in Section \ref{sec:results}.
Section \ref{sec:conclusion} concludes.
\ref{app:errors} details our error estimation procedure for the PES fits.
As a help to the reader, Table \ref{tab:acronyms} lists the acronyms
used frequently in this article.

\begin{table}
\centering
\begin{tabular}{ll}
\hline
MCTDH & Multi-Configuration Time-Dependent Hartree\\
ML-MCTDH & Multi-Layer MCTDH\\
PES & Potential Energy Surface\\
MGPF & Multi-Grid Potfit\\
MCPF & Monte-Carlo Potfit\\
MLPF & Multi-Layer Potfit\\
FG-MLPF & Full-Grid Multi-Layer Potfit\\
RS-MLPF & Random-Sampling Multi-Layer Potfit\\
SOP & Sum of Products\\
MLOp & Multi-Layer Operator\\
SPP & single-particle potential a.k.a. ``natural potential''\\
\hline
\end{tabular}
\caption{%
List of acronyms used frequently in this article.
\label{tab:acronyms}}
\end{table}

\section{Methodological Background}
\label{sec:methods}

\subsection{MCTDH, ML-MCTDH, and form of the Hamiltonian}
\label{sec:mctdh}

The multi-configuration time-dependent Hartree (MCTDH) method
\cite{mey90:73,bec00:1,mey09:book,mey12:351}
and its multi-layer variant (ML-MCTDH)
\cite{wan03:1289,man08:164116,ven11:044135}
have been widely discussed in detail elsewhere.
Here we only review a few key concepts of these methods,
as far as they are relevant for the present investigation.

MCTDH derives its efficiency from a compact representation of
the wavefunction, which reads
\begin{equation}
\Psi(q_1, \ldots, q_f, t) =
\sum_{j_1=1}^{n_1} \cdots \sum_{j_f=1}^{n_f}
A_{j_1 \cdots j_f}(t) \,
\varphi_{j_1}^{(1)}(q_1, t) \cdots \varphi_{j_f}^{(f)}(q_f, t)
\label{eq:mcdth}
\end{equation}
where $f$ denotes the number of degrees of freedom (DOFs) of the system,
$q_1$ through $q_f$ are the system coordinates, $t$ is the time,
$n_1$ through $n_f$ are expansion orders, and $A$ is a vector of coefficients
(the so-called ``$A$-vector'').
The ``single-particle functions'' (SPFs) $\varphi^{(\kappa)}_{j_\kappa}$ for
the $\kappa$-th DOF ($1 \leq \kappa \leq f$) form a time-dependent orthonormal set.
Hence \refeq{eq:mcdth} constitutes an expansion of $\Psi$ in a
time-dependent product basis, formed by products of the SPFs.
In turn, the SPFs are expressed in a time-independent ``primitive'' basis:
\begin{equation}
\varphi^{(\kappa)}_{j}(q_\kappa, t) =
\sum_{\alpha=1}^{N_\kappa}
A^{(\kappa)}_{j; \alpha}(t) \, \chi^{(\kappa)}_{\alpha}(q_\kappa)
\label{eq:spf}
\end{equation}
The primitive basis functions $\chi^{(\kappa)}_{\alpha}$ are usually
chosen to be the basis functions of a discrete variable representation (DVR)
or of other grid-based representations (\eg FFT), so that $N_\kappa$ denotes the
number of grid points for the $\kappa$-th DOF.
Note that a direct representation of $\Psi$ as an expansion in products of
primitive basis functions (\ie the ``standard method'') would require
$\prod_{\kappa=1}^f N_\kappa \approx N^f$ coefficients, while the MCTDH
representation only requires $\prod_{\kappa=1}^f n_\kappa \approx n^f$
coefficients for the $A$-vector plus $\sum_{\kappa=1}^f n_\kappa N_\kappa \approx fnN$
coefficients for the SPFs, cf.\ \refeq{eq:spf}. If $n \ll N$, this constitutes a strong
reduction of the storage requirements for the wavefunction.
This condition can usually be maintained during the time-evolution of $\Psi$,
at least for some time, because the equations of motion for MCTDH (\ie how the
coefficients $A$ and $A^{(\kappa)}$ evolve over time) are derived from the
Dirac-Frenkel variational principle \cite{dir30:376,fre34}, so that the set of SPFs
is variationally optimal for approximating the full $\Psi$ at each instant.

However, the number of coefficients still scales as $n^f$, \ie exponentially
with the number of DOFs.  A further reduction can be achieved via
``mode combination'', in which several DOFs are combined into one
large multi-dimensional mode.  Hence the SPFs are not one- but multi-dimensional:
\begin{eqnarray}
\Psi(Q_1, \ldots, Q_d, t) &=&
\sum_{j_1=1}^{n_1} \cdots \sum_{j_d=1}^{n_d}
A_{j_1 \cdots j_d}(t) \,
\varphi_{j_1}^{(1)}(Q_1, t) \cdots \varphi_{j_d}^{(d)}(Q_d, t)
\label{eq:mcdth-mc}
\\
\varphi^{(\kappa)}_{j}(Q_\kappa, t) &=& \varphi^{(\kappa)}_{j}(q_{a_\kappa}, \ldots, q_{b_\kappa}, t)
\nonumber \\
& = &
\sum_{\alpha_1=1}^{N_{a_\kappa}} \cdots \sum_{\alpha_p=1}^{N_{b_\kappa}}
A^{(\kappa)}_{j; \alpha_1 \cdots \alpha_p}(t) \,
\chi^{(a_\kappa)}_{\alpha_1}(q_{a_\kappa}) \cdots \chi^{(b_\kappa)}_{\alpha_p}(q_{b_\kappa})
\label{eq:spf-mc}
\end{eqnarray}
where $Q_\kappa = (q_{a_\kappa}, \ldots, q_{b_\kappa})$ denotes the multi-dimensional
coordinate for mode $\kappa$.  If the set of $f$ DOFs is combined into
$d$ modes of $p$ DOFs each (\ie $f=pd$), and if each mode employs
$n$ SPFs, then the number of coefficients needed to represent $\Psi$
is $n^d + d n N^p$.  This achieves another strong reduction in the
total number of coefficients.

Taken to the extreme, \ie combining the DOFs into only few (say, $d=2$) very large modes,
the $A$-vector becomes very small ($\sim n^d$) but the SPFs require
a large number of coefficients ($\sim d n N^{f/d}$).  At this stage it makes sense
to introduce another expansion layer, by decomposing the mode $Q_\kappa$ 
into a set of $p_\kappa$ sub-modes, $Q_\kappa = (Q_{\kappa,1}, \ldots, Q_{\kappa, p_\kappa})$,
where each sub-mode $Q_{\kappa,\lambda}$ is again a set of DOFs.
The mode-$\kappa$ SPFs are then expanded in products of sub-mode SPFs,
similar to \refeq{eq:mcdth}:
\begin{eqnarray}
\varphi^{(\kappa)}_{j}(Q_\kappa, t)
&=& \varphi^{(\kappa)}_{j}(Q_{\kappa,1}, \ldots, Q_{\kappa,p_\kappa}, t)
\nonumber \\
&=&
\sum_{k_1=1}^{n_{\kappa,1}} \cdots \sum_{k_{p_{\kappa}=1}}^{n_{\kappa,p_\kappa}}
A^{(\kappa)}_{j; k_1 \cdots k_{p_\kappa}}(t)
\prod_{\lambda=1}^{p_\kappa} \varphi^{(\kappa,\lambda)}_{k_\lambda}(Q_{\kappa,\lambda}, t)
\end{eqnarray}
In a two-layer scheme, the sub-mode SPFs $\varphi^{(\kappa,\lambda)}$ are again
expressed in products of primitive basis functions. But one may as well introduce
further layers of SPFs, which leads to a hierarchical expansion of $\Psi$ and gives
rise to the multi-layer MCTDH method (ML-MCTDH).
The total number of coefficients required to represent $\Psi$ in the ML-MCTDH
can be shown, under moderate assumptions, to scale only polynomially
instead of exponentially with the number of DOFs
(see \cite{man08:164116} and \cite{ott14:014106} for derivation and discussion).
As for the standard MCTDH method, the equations of motion for ML-MCTDH
can be derived using the Dirac-Frenkel variational principle.

In the present work, we are interested in computing the ground state of
the \HHHOO{} molecule.  A simple way to achieve this is by relaxation, \ie
by propagating some
initial state $\Psi_0$ in negative imaginary time $t=-i\tau$, for $\tau \to \infty$ \cite{kos86:223}.
This has the effect of exponentially damping all excited state components
present in $\Psi_0$ relative to the ground state component, so that the
propagated $\Psi$ converges to the ground state $\Psi_\text{GS}$ eventually.
However, convergence can be slow if there are low-lying excited states.
An alternative is the ``improved relaxation'' method
\cite{mey06:179, dor08:224109} where, in turn, the SPFs are relaxed and the
$A$-vector is obtained by diagonalization. This procedure is iterated until convergence is
reached. Not only does the improved relaxation method converge faster
to the ground state than regular relaxation, but a block version of the algorithm
is additionally able to compute excited states simultaneously.  Unfortunately, the Heidelberg MCTDH
package \cite{mlmctdh} currently supports (block) improved relaxation
calculations only for MCTDH wavefunctions but not for ML-MCTDH wavefunctions,
although it would certainly be possible to do so \cite{wan15:7951}.
In the present work, we therefore resorted to regular relaxation for our
ML-MCTDH calculations.

A requirement for an efficient operation of (ML-)MCTDH is that the Hamiltonian
must be expressed in a form which allows the efficient evaluation of high-dimensional
matrix elements and related quantities, which are encountered during the evaluation
of the equations of motion. For example, the expectation value
$\langle \Psi | \hat{H} | \Psi \rangle$ formally constitutes an $f$-dimensional
integral, but it can be evaluated efficiently if $\hat{H}$ has a representation
as a sum of products (SOP) of one-dimensional operators (here assuming for
simplicity that mode combination is not used), \ie
\begin{equation}
\hat{H} = \sum_{r=1}^s c_r \prod_{\kappa=1}^f \hat{h}_r^{(\kappa)}
\label{eq:sop}
\end{equation}
where the operators $\hat{h}_r^{(\kappa)}$ operate on the DOF $q_\kappa$ only.
Then the expectation value is evaluated as (cf.\ \refeq{eq:mcdth})
\begin{equation}
\langle \Psi | \hat{H} | \Psi \rangle =
\sum_r c_r
\sum_{j_1 \cdots j_f} A^{*}_{j_1 \cdots j_f}
\sum_{k_1 \cdots k_f} A_{k_1 \cdots k_f}
\prod_{\kappa=1}^f
\langle \varphi^{(\kappa)}_{j_\kappa} | \hat{h}^{(\kappa)}_r | \varphi^{(\kappa)}_{k_\kappa} \rangle
\label{eq:sop-xp}
\end{equation}
which boils down to a series of one-dimensional integrals, a sequential
series of tensor-matrix products, and a final contraction over $j_1 \cdots j_f$, all of
which can be performed relatively efficiently unless the $A$-vector is too large.
These steps need to be carried out for each term $r$ in the SOP expansion, \refeq{eq:sop}.
Hence the computational effort for evaluating \refeq{eq:sop-xp}, as well
as for similar terms appearing in the MCTDH equations of motion, scales linearly with $s$,
\ie with the length of the SOP expansion.

The kinetic energy operator (KEO) often naturally has SOP form,
at least if a suitable coordinate system is chosen, like polyspherical coordinates \cite{gat09:1}.
On the other hand, for non-model systems the potential energy operator usually doesn't
come in SOP form, but is given by a multi-dimensional potential energy surface (PES) which
is obtained as a fit to a large set of electronic structure calculations
on a set of judiciously chosen geometries.
Such a PES must be transformed to SOP form before it can be used with MCTDH,
as will be discussed in Section \ref{sec:pesfit}.

For ML-MCTDH, the SOP form can also be evaluated efficiently \cite{man08:164116,ven11:044135}.
However, as ML-MCTDH can handle much larger systems (\ie with more DOFs) than MCTDH,
the length of the SOP expansion can become a problematic issue.
In Ref.\ \cite{ott14:014106}, an alternative representation of the Hamiltonian as
a ``multi-layer operator'' (MLOp) was developed, which possesses the same
compact hierarchical structure as the ML-MCTDH wavefunction.
It was demonstrated that this structure can lead to a much more efficient
evaluation of the terms in the ML-MCTDH equations of motion, already
for systems as small as $f=4$, and with potentially vast computational savings
for large systems.  As for the SOP form, a way to transform general PES into
MLOp form is needed, which will likewise be discussed in Section \ref{sec:pesfit}.

\subsection{Variational grid-based PES Fitting Methods}
\label{sec:pesfit}

One popular method for transforming a PES $V(q_1, \ldots, q_f)$ into SOP form
is the Potfit algorithm \cite{jae96:7974}.
This grid-based method starts by evaluating the PES on a product grid
of points $(q^{(\kappa)}_{\alpha_\kappa}, \kappa=1 \ldots f)$, which
yields an $f$-dimensional tensor
\begin{equation}
V_{\alpha_1 \cdots \alpha_f} = V(q^{(1)}_{\alpha_1}, \ldots, q^{(f)}_{\alpha_f})
\quad ; \quad 1 \leq \alpha_\kappa \leq N_\kappa
\end{equation}
Similar to the MCTDH ansatz \refeq{eq:mcdth}, this tensor is then approximated
by expanding it in a basis of ``single-particle potentials'' (SPPs)
\begin{equation}
V_{\alpha_1 \cdots \alpha_f} \approx
\tilde{V}_{\alpha_1 \cdots \alpha_f} =
\sum_{j_1=1}^{m_1} \cdots \sum_{j_f=1}^{m_f} C_{j_1 \cdots j_f} \,
v^{(1)}_{\alpha_1 j_1} \cdots v^{(f)}_{\alpha_f j_f}
\label{eq:potfit}
\end{equation}
where the $v^{(\kappa)}_{\alpha_\kappa j_\kappa}$
are the components of the SPPs for DOF $\kappa$,
$C$ is an $f$-dimensional coefficient tensor (the ``core tensor''),
and the $m_\kappa$ are expansion orders.
\refeq{eq:potfit} is also known as a (truncated) Tucker decomposition \cite{tuc63:122}.

To obtain the SPPs, Potfit builds a ``potential density matrix''
$\vct{\rho}^{(\kappa)} \in \mathds{R}^{N_\kappa \times N_\kappa}$
for each DOF $\kappa$ by contracting the
tensor $V$ with itself over all DOFs except the $\kappa$-th one, \ie
\begin{equation}
\rho^{(\kappa)}_{\alpha \beta} =
\sum_{\alpha_1=1}^{N_1} \cdots \sum_{\alpha_{\kappa-1}=1}^{N_{\kappa-1}}
\sum_{\alpha_{\kappa+1}=1}^{N_{\kappa+1}} \sum_{\alpha_f=1}^{N_f}
V_{\alpha_1 \cdots \alpha_{\kappa-1} \alpha \alpha_{\kappa+1} \cdots \alpha_f}
V_{\alpha_1 \cdots \alpha_{\kappa-1} \beta\alpha_{\kappa+1} \cdots \alpha_f}
\label{eq:potrho}
\end{equation}
These density matrices are then diagonalized to obtain eigenvalues
$\lambda^{(\kappa)}_j$ (the ``natural weights'') and eigenvectors
$\vct{v}^{(\kappa)}_j$ (the ``natural potentials''), $1 \leq j \leq N_\kappa$.
Only the dominant $m_\kappa$ natural potentials (\ie those corresponding to the largest
natural weights) are kept as the SPPs.  Finally, the core tensor $C$ is found
by projecting the tensor $V$ onto products of the SPPs.
A mathematically equivalent algorithm is known as the higher-order
singular value decomposition (HOSVD) \cite{lat00:1253}.

Except for $f=2$, Potfit does not yield the optimal
SPPs and core tensor. However, the error
$\Delta = \| V - \tilde{V }\|$  introduced by Potfit
can be bounded by the sum of the neglected
natural weights \cite{pel13:014108},
\begin{equation}
\frac{1}{f-1} \sum_{\kappa=1}^{f} \sum_{j>m_\kappa} \lambda^{(\kappa)}_j
\leq \Delta_\text{opt}^2 \leq \Delta^2 \leq
\sum_{\kappa=1}^{f} \sum_{j>m_\kappa} \lambda^{(\kappa)}_j
\label{eq:potfit-error}
\end{equation}
where $\Delta_\text{opt}$ is the best possible approximation of $V$
in the form \refeq{eq:potfit} with fixed expansion orders $m_\kappa$.
This shows that the Potfit approximation is at most a factor of
$\sqrt{f-1}$ worse than the best possible approximation of the same form and size.
It also shows that in order to significantly increase the accuracy of the
approximation, it will be necessary to increase the expansion orders $m_\kappa$,
as this is the only way to reduce the lower bound in \refeq{eq:potfit-error}.

The number of SOP terms generated by the Potfit expansion
according to \refeq{eq:potfit} is the number of entries in the core tensor, $\sim m^f$.
This can be reduced by a factor of $m$ by contracting the core tensor with the
SPPs of one DOF (the ``contracted mode''), \eg the first one:
\begin{eqnarray}
D^{(1)}_{\alpha_1 j_2 \cdots j_f} & = & \sum_{j_1=1}^{m_1} C_{j_1 \cdots j_f} v^{(1)}_{\alpha_1 j_1}
\label{eq:contraction}
\\
\tilde{V}_{\alpha_1 \cdots \alpha_f} & = &
\sum_{j_2=1}^{m_2} \cdots \sum_{j_f=1}^{m_f} D^{(1)}_{\alpha_1 j_2 \cdots j_f} \,
v^{(2)}_{\alpha_2 j_2} \cdots v^{(f)}_{\alpha_f j_f}
\label{eq:potfit-contracted}
\end{eqnarray}
In doing so, one may as well choose $m_1=N_1$ in order to reduce the Potfit error.
The resulting expansion, \refeq{eq:potfit-contracted}, thus generates $\sim m^{f-1}$ SOP terms.

A major disadvantage of Potfit is that it needs to build the PES tensor $V$ on the
full product grid, which has $\sim N^f$ entries.  In practice this limits the applicability
of Potfit to systems with 6--8 DOFs, otherwise the computation of all the entries of $V$
and of the potential density matrices becomes too expensive.
Recently, two methods have been proposed for overcoming this limitation of Potfit,
\ie they avoid running over the full product grid.
Reviewing \refeq{eq:potrho}, we note that the potential density matrices
for DOF $\kappa$ are computed by integrating over the \emph{complementary} grid,
\ie the product grid of all DOFs except the $\kappa$-th.
Instead of performing this $(f-1)$-dimensional integration exactly, one
can perform it approximately.
In \emph{Monte-Carlo Potfit} (MCPF) \cite{mcpf}, the integration over the
complementary grid is replaced by a Monte-Carlo integration, \ie by sampling the
grid points for integration either uniformly from the full grid or via the
Metropolis-Hastings algorithm \cite{met53:1087,has70:97}. 
Additional steps are then required for computing the core tensor, to correct for the error introduced
by approximating the integration over the complementary grid.
In \emph{multigrid Potfit} (MGPF) \cite{pel13:014108},
in addition to the original full ``fine'' product grid, a ``coarse'' product grid is introduced,
which consists of a subset of the fine grid points.
In between these two limiting cases one then defines a series of partial grids
where one mode lies on the fine grid and the rest lie on the coarse grid.
The SPPs on the fine grid are obtained through an interpolation process in which SPPs from a coarse-grid Potfit
are transferred onto the fine grid by multiplication with a product of approximated density matrices
computed on the partial grids.
For details of both MGPF and MCPF the reader is referred to the original publications
\cite{pel13:014108, mcpf}.

Another problem of Potfit is that the resulting number of SOP terms ($\sim m^{f-1}$, cf.\ \refeq{eq:potfit-contracted})
grows exponentially with the number $f$ of DOFs.  This problem is not overcome by
MGPF or MCPF, as they ``only'' provide more efficient methods to compute a
representation in SOP form, but cannot overcome the fundamental limitation
given by the Potfit error estimate, \refeq{eq:potfit-error}.
Especially if a high accuracy of the PES representation is needed, it is necessary
to increase the expansion orders $m_\kappa$, so that the number of terms in
the Potfit expansion quickly becomes prohibitive due to its exponential scaling.
In the context of ML-MCTDH, this fundamental problem can be overcome by
abandoning the SOP form and switching to the hierarchical MLOp form for
representing the PES. Again, this poses the problem of how to transform
a general PES into MLOp form, and the \emph{multi-layer Potfit} (MLPF) method \cite{ott14:014106}
has been developed for this purpose, based on 
the hierarchical singular value decomposition \cite{grasedyck_hierarchical_2010}.
In Ref.\ \cite{ott14:014106}, MLPF was applied to a relatively small system (diatom-diatom
scattering) and the PES which had to be transformed was 5-dimensional. Despite the
small dimensionality, MLPF achieved a strong reduction of the computational
resources required for the ML-MCTDH propagations (both in CPU and RAM),
compared to using Potfit at the same accuracy. 

As originally conceived, MLPF suffers from the same problem as Potfit,
in that it requires running over the full product grid in order to transform the
PES into MLOp form, which again limits applicability to 6--8 DOFs.
In the present article, we will explore two options for overcoming this limitation.
The first option makes use of the fact that the first step of MLPF is identical
to Potfit, except that contraction is not used.  Hence one may use MGPF or MCPF
to compute an initial Potfit, undo the contraction\footnote{%
The potential density matrix for the contracted mode can be computed
(approximately) by self-contracting the $D$-tensor, cf.\ \refeq{eq:contraction},
to form $\rho^{(1)}_{\alpha \beta}$.} to obtain SPPs for the contracted
mode as well as the core tensor $C$ (\cf \refeq{eq:potfit}),
and then perform the remaining steps of MLPF.  These steps are
computationally cheap compared to the initial Potfit, as they don't need
to operate on the full-grid PES but only on the $C$ tensor which is
much smaller. MLPF reshapes the $C$ tensor by combining low-dimensional
modes into higher-dimensional modes and performs another HOSVD step in order to
reduce its data, and repeats this process until only a few (usually 2 or 3)
large-dimensional modes are left.
In contrast to performing MLPF on the full-grid PES, this way of using MLPF
as a postprocessing option to MGPF/MCPF fully avoids the full-grid problem,
and is rather fast and simple to perform.
However, the accuracy of this approach can be
limited, because in a high-accuracy setting the initial Potfit produced
by MGPF/MCPF might become very large due to its inherent scaling behavior.

The second option is to integrate the approximation ideas from MGPF or MCPF
directly into MLPF. The goal is to avoid the explicit computation of the core tensor,
which easily becomes the bottleneck if high accuracy (\ie large expansion orders
$m_\kappa$) are required. As a first step in this direction, in Section \ref{sec:rs-mlpf}
we will present a variant of MLPF which avoids running over the full product grid
by using uniform random sampling as in MCPF, while also avoiding to compute
the core tensor as an intermediate step. As we will show, this new variant of MLPF
is able to produce rather accurate PES fits in MLOp form with modest computational
resources, however its implementation currently only has prototype status so that
it may not be readily applicable for systems with a different number of modes or
a different multi-layer structure than investigated here.
To make the distinction between the new variant and the original MLPF method
more clear, we will henceforth refer to the original method as ``full-grid MLPF'' (FG-MLPF).

\subsection{Random Sampling Multi-Layer Potfit (RS-MLPF)}
\label{sec:rs-mlpf}

In order to avoid an overly technical exposition, here we present
the RS-MLPF algorithm for a balanced binary multi-layer tree
with eight uncombined primitive modes, \ie a three-layer tree.
The generalization to arbitrary tree structures is relatively straight-forward,
but would require an overabundance of additional notation.
In the following, $\kappa = 1 \ldots 8 $ numbers the DOFs,
$N_\kappa$ denotes the number of grid points for the $\kappa$-th DOF,
and $\alpha_\kappa = 1 \ldots N_\kappa$ indexes these grid points.

In the first step, RS-MLPF proceeds like Monte Carlo Potfit \cite{mcpf}.
The potential density matrix for the 1st DOF is defined as
\begin{equation}
\rho^{(1)}_{\alpha_1 \alpha'_1} = \sum_{\alpha_2 \cdots \alpha_8} V_{\alpha_1 \alpha_2 \cdots \alpha_8} V_{\alpha'_1 a_2 \cdots \alpha_8}
\end{equation}
where $V$ denotes the eight-dimensional PES tensor defined on the full grid.
This 7D integration is then replaced by a Monte Carlo integration, using
$R$ sampling points $\{(\alpha^r_2 \cdots \alpha^r_8) | r=1 \ldots R\}$ :
\begin{equation}
\rho^{(1)}_{\alpha_1 \alpha'_1} = \sum_{r} V_{\alpha_1 \alpha^r_2 \cdots \alpha^r_8} V_{\alpha'_1 \alpha^r_2 \cdots \alpha^r_8}
\end{equation}
To avoid a rank-deficit $\rho^{(1)}$, one should choose $R>N_1$.
In practice, we choose $R=q N_1$, where $q$ is the \emph{oversampling parameter}.
A larger $q$ will result in using more sampling points, and thus in a more
accurate estimation of the potential density matrices.
Despite the Monte Carlo approximation, $\rho^{(1)}$  is symmetric and positive semi-definite. Diagonalizing $\rho^{(1)}$
yields natural weights and natural potentials, of which only the dominant $m_1$ ones are kept.
Mathematically equivalent, we reshape the sampled $V$ tensor into a matrix $V^{(1)}$ with entries
$V^{(1)}_{\alpha_1 r} = V_{\alpha_1 \alpha^r_2 \cdots \alpha^r_8}$, and perform a singular value decomposition (SVD)
on this matrix. Its left singular vectors then yield the natural potentials, and the squares of the singular
values yield the natural weights. Numerically this approach is more stable, hence it is the choice used in practice.
Using the same procedure for the other DOFs, we obtain all natural potentials for the lowest
(third, in our example) layer:
\begin{equation}
u^{(\kappa)}_{\alpha_\kappa i_\kappa} \quad;\: \kappa=1 \ldots 8,\: \alpha_\kappa=1 \ldots N_\kappa,\: i_\kappa=1 \ldots m_\kappa
\end{equation}
The core tensor for these natural potentials is
\begin{equation}
C_{i_1 \cdots i_8} = \sum_{\alpha_1 \cdots \alpha_8} u^{(1)}_{\alpha_1 i_1} \cdots u^{(8)}_{\alpha_8 i_8} V_{\alpha_1 \cdots \alpha_8}
\label{eq:rsmlpf-C}
\end{equation}
but we can actually avoid computing it, because in MLPF $C$ is only  neeeded to compute
the density matrices for the next higher layer. Introducing the mode-combined indices
\begin{equation}
b_1=(i_1 i_2) \quad b_2=(i_3 i_4) \quad b_3=(i_5 i_6) \quad b_4=(i_7 i_8)
\quad,
\end{equation}
the density matrix for mode 1 in the second layer (which comprises DOFs 1 and 2) reads
\begin{equation}
\bar{\rho}^{(1)}_{b_1 b'_1} = \sum_{b_2 b_3 b_4} C_{b_1 b_2 b_3 b_4} C_{b'_1 b_2 b_3 b_4}
\end{equation}
We can now formally insert the expression for $C$, \refeq{eq:rsmlpf-C}, and for the purpose
of this computation we can furthermore choose full expansion orders 
$m_3=N_3, \ldots, m_8=N_8$ which makes the natural potential bases for DOFs 3 through 8 complete.
Using this completeness property, we find that $\bar{\rho}^{(1)}$ can be computed via
\begin{align}
\bar{\rho}^{(1)}_{b_1 b'_1} &= \sum_{\alpha_3 \cdots \alpha_8} D^{(1)}_{b_1 \alpha_3 \cdots \alpha_8}
D^{(1)}_{b'_1 \alpha_3 \cdots \alpha_8} \\
D^{(1)}_{b_1 \alpha_3 \cdots \alpha_8} &= \sum_{\alpha_1 \alpha_2} u^{(1)}_{\alpha_1 i_1} u^{(2)}_{\alpha_2 i_2}
V_{\alpha_1 \cdots \alpha_8}
\label{eq:rsmlpf-D}
\end{align}
Here, \(\bar{\rho}^{(1)}\) is computed via a 6D integration, which we again replace by
a Monte Carlo integration, using a new set of $\bar{R}$ sampling points
\(\{ A^r := (\alpha^r_3 \cdots \alpha^r_8)|r=1 \ldots \bar{R}\}\):
\begin{equation}
\bar{\rho}^{(1)}_{b_1 b'_1} = \sum_r D^{(1)}_{b_1 \alpha^r_3 \cdots \alpha^r_8} D^{(1)}_{b'_1 \alpha^r_3 \cdots \alpha^r_8} 
\quad.
\label{eq:rsmlpf-rhobar}
\end{equation}
Again, to avoid rank deficiency one should choose \(\bar{R}>m_1 m_2\) and in practice we use $\bar{R}=q m_1 m_2$.
From expression (\ref{eq:rsmlpf-rhobar}) it becomes clear that \(D^{(1)}_{bA}\) doesn't need to be known on the full
product grid, but only on the sampling points \(\{A^r\}\). 
It now remains to evaluate \(D^{(1)}_{bA}\) for these points. In principle, this could be achieved by
directly evaluating \refeq{eq:rsmlpf-D}, but this may become too costly for large modes.
Alternatively, we note that the PES approximation is now given by
\begin{equation}
\tilde{V}_{\alpha_1 \cdots \alpha_8} =
\sum_{i_1 i_2} D^{(1)}_{i_1 i_2 \alpha_3 \cdots \alpha_8} u^{(1)}_{i_1 \alpha_1} u^{(2)}_{i_2 \alpha_2}
\end{equation}
and that the optimal approximation is achieved by minimizing \(\|V-\tilde{V}\|^2\) w.r.t. \(D^{(1)}_{bA}\).
This leads to the equation
\begin{align}
&\sum_{\alpha_1 \alpha_2} \Omega^{(1)}_{\alpha_1 \alpha_2, b} V_{\alpha_1 \alpha_2 A}
= \sum_{b'} \left( \sum_{\alpha_1 \alpha_2} \Omega^{(1)}_{\alpha_1 \alpha_2,b} \Omega^{(1)}_{\alpha_1 \alpha_2,b'} \right)
D^{(1)}_{b' A}
\\
&\text{where}\quad \Omega^{(1)}_{\alpha_1 \alpha_2, b} = u^{(1)}_{\alpha_1 i_1} u^{(2)}_{\alpha_2 i_2}
\end{align}
If the summation over \(\alpha_1 \alpha_2\) is carried out fully, then the term in parentheses yields \(\delta_{bb'}\) (due to orthonormality of the natural potentials) and one arrives back at \refeq{eq:rsmlpf-D}. Instead we can replace this sum (both on the LHS and on the RHS) by a Monte Carlo integration. Using the new set of \(\bar{S}\) sampling points \(\{(\alpha^s_1 \alpha^s_2)|s=1 \ldots \bar{S}\}\), this results in
\begin{align}
&D^{(1)}_{bA} = \sum_{b'} (X^{(1)})^{-1}_{bb'} \sum_s \Omega^{(1)}_{\alpha^s_1 \alpha^s_2, b} V_{\alpha^s_1 \alpha^s_2 A}
\\
&\text{where}\quad X^{(1)}_{bb'} = \sum_s \Omega^{(1)}_{\alpha^s_1 \alpha^s_2,b} \Omega^{(1)}_{\alpha^s_1 \alpha^s_2,b'}
\end{align}
Here we must choose \(\bar{S}>m_1 m_2\), otherwise the matrix $X^{(1)}$ won't be invertible.
Again, our choice in practice is $\bar{S} = q m_1 m_2$.
Once \(D^{(1)}_{bA}\) is computed, the density matrix \(\bar{\rho}^{(1)}\) is easily computed 
via \refeq{eq:rsmlpf-rhobar} and diagonalized (or mathematically equivalent,
an SVD of the matrix \(D^{(1)}_{bA}\) is performed, similar to the previous layer).
Doing so for all four modes of layer 2 yields their natural potentials,
and again, only the dominant ones are kept:
\begin{equation}
\bar{u}^{(\kappa)}_{b_\kappa j_\kappa}\quad;
\kappa=1\ldots 4,\: b_\kappa=1\ldots m^2,\: j_\kappa=1\ldots \bar{m}_\kappa
\end{equation}
Thus we have achieved the computation of the natural potentials for layer 2,
while avoiding to compute the intermediate core tensor $C$.

If we stopped the expansion at layer 2, we would need to use the core tensor
\begin{equation}
\bar{C}_{j_1 \cdots j_4} = \sum_{b_1 \cdots b_4}
\bar{u}^{(1)}_{b_1 j_1} \cdots \bar{u}^{(4)}_{b_4 j_4} C_{b_1 \cdots b_4}
\label{eq:rsmlpf-Cbar}
\end{equation}
Considering that we don't actually know $C$, we again aim to avoid this computation.
To this end we  formally insert the definition of $C$, \refeq{eq:rsmlpf-C}, into \refeq{eq:rsmlpf-Cbar}:
\begin{align}
\bar{C}_{j_1 \cdots j_4}
&= \sum_{i_1 \cdots i_8} \bar{u}^{(1)}_{i_1 i_2 j_1} \cdots \bar{u}^{(4)}_{i_7 i_8 j_4}
\sum_{\alpha_1 \cdots \alpha_8} u^{(1)}_{\alpha_1 i_1} u^{(2)}_{\alpha_2 i_2} \cdots u^{(7)}_{\alpha_7 i_7} u^{(8)}_{\alpha_8 i_8}
V_{\alpha_1 \cdots \alpha_8}
\nonumber\\
&= \sum_{\alpha_1 \cdots \alpha_8}
\left[ \sum_{i_1 i_2} u^{(1)}_{\alpha_1 i_1} u^{(2)}_{\alpha_2 i_2} \bar{u}^{(1)}_{i_1 i_2 j_1} \right]
\times \cdots \times
\left[ \sum_{i_7 i_8} u^{(7)}_{\alpha_7 i_7} u^{(8)}_{\alpha_8 i_8} \bar{u}^{(4)}_{i_7 i_8 j_4} \right]
V_{\alpha_1 \cdots \alpha_8}
\nonumber\\
&= \sum_{\alpha_1 \cdots \alpha_8}
\tilde{u}^{(1)}_{\alpha_1 \alpha_2 j_1} \cdots \tilde{u}^{(4)}_{\alpha_7 \alpha_8 j_4}
V_{\alpha_1 \cdots \alpha_8}
\end{align}
where we have introduced the abbreviations
\begin{equation}
\tilde{u}^{(1)}_{\alpha_1 \alpha_2 j_1} = \sum_{i_1 i_2} u^{(1)}_{\alpha_1 i_1} u^{(2)}_{\alpha_2 i_2} \bar{u}^{(1)}_{i_1 i_2 j_1}
\label{eq:utilde}
\end{equation}
etc.  At this level, the approximation for the PES tensor $V$ reads
\begin{equation}
\tilde{V}_{\alpha_1 \cdots \alpha_8} =
\sum_{j_1 \cdots j_4} \bar{C}_{j_1 \cdots j_4} \tilde{u}^{(1)}_{\alpha_1 \alpha_2 j_1} \cdots \tilde{u}^{(4)}_{\alpha_7 \alpha_8 j_4}
\end{equation}
and as for the previous layer, the optimal $\bar{C}$ is obtained from minimizing
$\| V - \tilde{V} \|^2$ w.r.t. $\bar{C}_{J}$, where we have introduced the multi-index
$J = (j_1 \cdots j_4)$.
This eventually leads to the equation
\begin{align}
\bar{C}_{J} &= \sum_{J'} (\tilde{X}^{(-1)})_{J J'}
\sum_{A} \tilde{\Omega}_{A J'} V_{A}
\label{eq:cbar}
\\
\text{where}\quad
\tilde{X}_{J J'} &= \sum_{A} \tilde{\Omega}_{A J} \tilde{\Omega}_{A J'}
\label{eq:xtilde}
\\
\text{and}\quad
\tilde{\Omega}_{A J} &= \tilde{u}^{(1)}_{\alpha_1 \alpha_2 j_1} \cdots \tilde{u}^{(4)}_{\alpha_7 \alpha_8 j_4}
\label{eq:omegatilde}
\end{align}
and $A=(\alpha_1 \cdots \alpha_8)$ is a full-grid multi-index.
As before, the summations over $A$ are replaced by Monte Carlo integrations,
but to avoid dealing with a very large $\tilde{\Omega}$ matrix, we choose
a set of sampling points separately for the first four and the last four DOFS,
\ie we use the point set $\{A^{rs} = (\alpha^r_1 \cdots \alpha^r_4 \alpha^s_5 \cdots \alpha^s_8) \,|\,
r=1\ldots\bar{\bar{R}}, s=1\ldots\bar{\bar{S}}\}$.
This results in a separable structure of the matrices $\tilde{\Omega}$
and $\tilde{X}$ which greatly reduces the computational effort for the
linear algebra parts of Eqs.\ (\ref{eq:cbar})--(\ref{eq:omegatilde}) but
requires more PES evaluations.
To ensure invertibility of $\tilde{X}$, we must choose $\bar{\bar{R}} > \bar{m}_1 \bar{m}_2$
and $\bar{\bar{S}} > \bar{m}_3 \bar{m}_4$.
Starting with the already known natural potentials for layers 3 and 2, we can
evaluate Eqs.\ (\ref{eq:utilde}), (\ref{eq:omegatilde}), (\ref{eq:xtilde}), and (\ref{eq:cbar})
in sequence to obtain $\bar{C}$.

Finally, we introduce a next layer of mode-combined indices
\begin{equation}
c_1=(j_1 j_2) \quad c_2=(j_3 j_4)
\end{equation}
and rewrite the tensor $\bar{C}_{j_1 j_2 j_3 j_4}$ as a matrix $\bar{C}_{c_1 c_2}$,
on which we perform a singular value decomposition:
\begin{equation}
\bar{C}_{c_1 c_2} = \sum_{k_1 k_2} \bar{\bar{C}}_{k_1 k_2} \bar{\bar{u}}^{(1)}_{c_1 k_1} \bar{\bar{u}}^{(2)}_{c_2 k_2}
\end{equation}
where the top-level (L1) core tensor $\bar{\bar{C}}$ is diagonal.  The left ($\bar{\bar{u}}^{(1)}$)
and right ($\bar{\bar{u}}^{(2)}$) singular vectors
serve as the natural potentials for the two L1 modes,
and we only keep the $\bar{\bar{m}}$ dominant ones.
This completes the RS-MLPF algorithm, as we have now obtained the natural potentials
for all modes of all layers as well as the top-level core tensor.

For the purposes of this study, we have created a prototype implementation
of the RS-MLPF algorithm in the high-level programming language Julia \cite{Julia}.
Our implementation is currently
restricted to a balanced binary tree with two or three layers (\ie 4 or 8 primitive modes).
Moreover, all Monte Carlo integrations are simply done by uniform random sampling over all grid points.
Like for the FG-MLPF algorithm described in \cite{ott14:014106}, 
the algorithm automatically determines all the expansion orders
$m_1 \cdots m_8$, $\bar{m}_1 \cdots \bar{m}_4$, and $\bar{\bar{m}}$
by taking a user-prescribed threshold for the global RMS error and
distributing it evenly over all modes in the multi-layer tree.  However, the
errors introduced by the various Monte Carlo integrations are not taken into
account in this error control scheme, so that the actual fit error is larger
than for FG-MLPF.  This additional random sampling error (``RS error'')
can be reduced by increasing the oversampling parameter $q$.
We point out that our RS-MLPF implementation is not yet optimized nor parallelized,
but as we will show in Section \ref{sec:results}, it already enables us to obtain
highly accurate PES fits with rather modest computational resources.

\section{The \HHHOO{} Ion}
\label{sec:h3o2}

The \HHHOO{} ion is an interesting prototype for understanding
the anomalous high mobility of  the hydroxide ion in water
\cite{tuckerman_quantum_1997, tuckerman_nature_2002, huang_quantum_2004, diken_fundamental_2005, yu_rigorous_2006}.
Electronic structure calculations reveal that at its energetic minimum, \HHHOO{} posseses
a slighly asymmetric structure, where the bridging hydrogen is closer to one oxygen than
to the other \cite{huang_quantum_2004}.
However, quantum dynamical calculations show that the bridging hydrogen is
well shared between the two OH moieties, as there is a very low barrier
($\sim 70\,\ic$) for the proton transfer \cite{mccoy_quantum_2005}.

Our (ML-)MCTDH calculations make use of the PES3C
potential energy surface \cite{huang_quantum_2004, mccoy_quantum_2005},
which was constructed as a least-squares fit to $\sim$\ 23,000 energy points calculated
at the CCSD(T) level with an aug-cc-pVTZ basis set.
This PES is a more global variant of its predecessor PES2 \cite{huang_quantum_2004}, as it includes
a number of dissociative conformations.
The fit sports a relatively low root-mean-square (RMS) error of 18.0 and 103\ $\ic$ for energies
up to 6,000 and 30,000\ $\ic$, respectively.
An important feature of the PES3C surface is its permutational invariance, \ie
the PES value is invariant under interchange of like atoms (H with H, or O with O).

Even in its ground state, the \HHHOO{} ion is affected by large-amplitude motions
and strong anharmonicity, as the PES features two shallow double-well potentials:
one along the proton-transfer coordinate (with a barrier of 74\ $\ic$ according to PES3C),
and another along the torsional coordinate around the O-O axis (barriers of
374\ $\ic$ and 147\ $\ic$ at $0\dg$ and $180\dg$, respectively).
These features lead to strong correlations between the DOFs, which justify the need for
full-dimensional quantum calculations to investigate this system.

The PES3C surface has been previously used to compute the ground state
(zero-point) energy of \HHHOO{} as well as vibrationally excited states.
McCoy et al. \cite{mccoy_quantum_2005} report results from vibrational
configuration-interaction (VCI) and diffusion Monte-Carlo (DMC) calculations,
yielding ground state (GS) energies of $6625\ \ic$ (VCI) and $6605 \pm 5\ \ic$ (DMC),
respectively. However, these reported values were obtained with the PES2
surface, though further calculations with the PES3C surface differed by less
than $5\ \ic$ \cite{mccoy_quantum_2005} for the DMC results.
Additional results for the PES3C surface were obtained by Yu \cite{yu_rigorous_2006}
who used a two-layer Lanczos algorithm with mixed grid/non-direct product basis set
and reported a GS energy of $6623.5\ \ic$.
More recently, in a series of papers \cite{pel13:014108, pel14:42, pel17:100} Peláez et al. investigated ground and
excited states of \HHHOO{} with the MCTDH method, where improved relaxation was used to obtain the states
and MGPF was used to transform the PES3C surface into SOP form. For the most accurate PES fit,
the reported GS energy was $6602\ \ic$.

\begin{figure}[tb]
\center
\includegraphics[width=8cm]{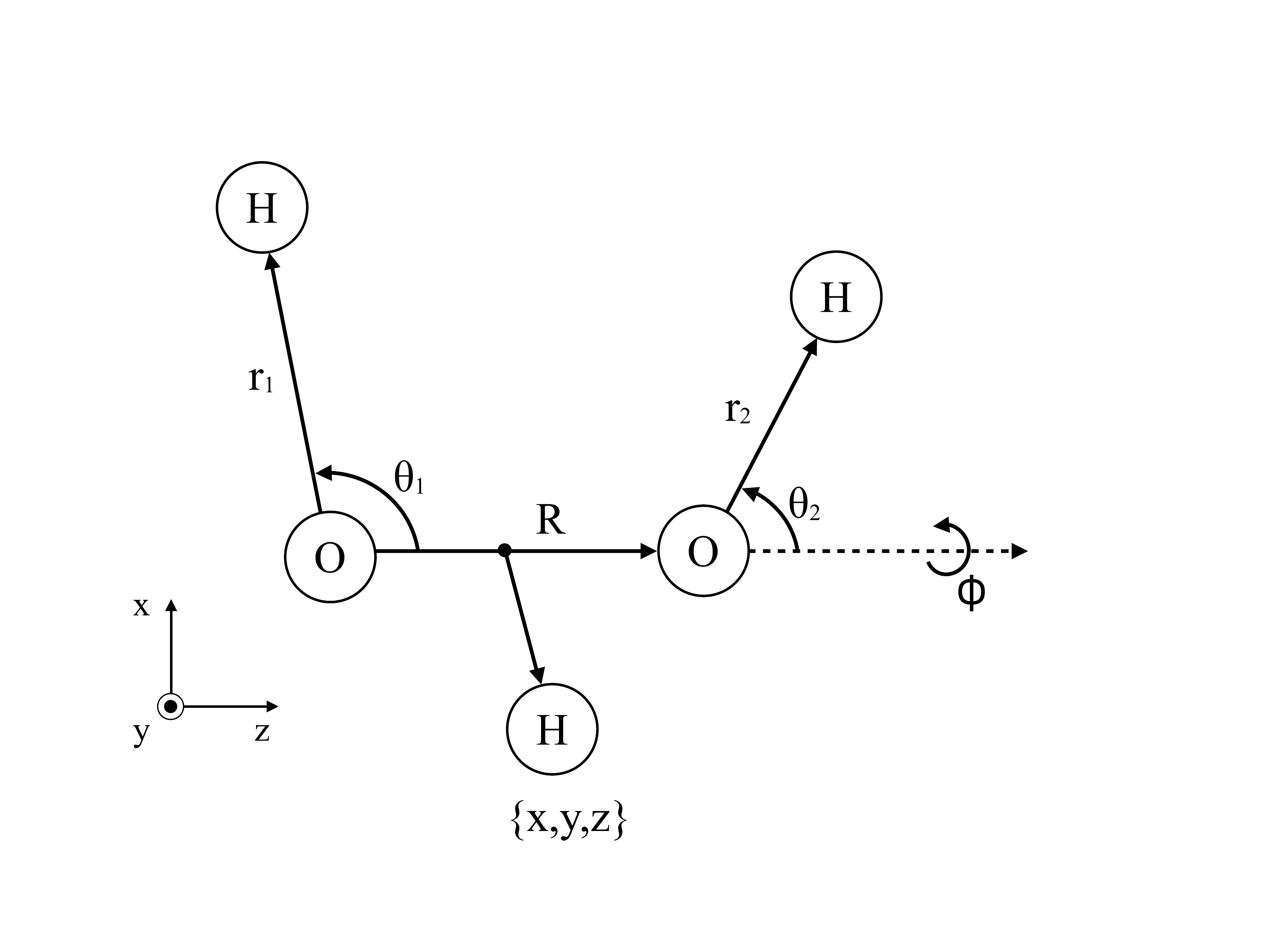}
\caption{Valence coordinates for \HHHOO.
\label{fig:h3o2}}
\end{figure}

In this work we use the same valence coordinate system as in the previous
works of Peláez et al. \cite{pel14:42, pel17:100}.
(cf.\ Fig.\ \ref{fig:h3o2}).
This coordinate system is defined by two vectors $\vct{r}_1$ and $\vct{r}_2$ for the two O-H moieties,
a vector $\vct{R}$ connecting the O atoms, and a vector $\vct{r}_3 = (x,y,z)$
connecting the center of the O-O vector $\vct{R}$ to the bridging H atom.
Ignoring the overall rotation, this yields nine coordinates for describing the system:
the O-H bond lengths $r_1$ and $r_2$, the O-O distance $R$, the position of the bridging H
$(x,y,z)$, the azimuthal angles $\theta_1$ and $\theta_2$ between the O-H vectors and the
O-O axis (though note that the calculations actually use $u_i = \cos \theta_i$,
cf.\ Table \ref{tab:dvr}), and the torsional angle $\phi$.
As suggested by Vendrell et al. \cite{ven09:234305}, we replace the coordinate $z$ by
a dimensionless variable $z_\text{red} = z / ( R - 2 d_0 )$ which can take values
in the range $[-0.5, 0.5]$. $d_0$ is a parameter which signifies the minimum allowed distance
of the bridging H to one of the O atoms. Here we use $d_0 = 1.6\, a_0$.
In this way, highly energetic conformations in which the bridging H lies too close to one of
the O atoms are avoided.

As noted by Yu \cite{yu_rigorous_2006} and Peláez et al. \cite{pel13:014108},
the PES3C energy surface exhibits ``holes'', i.e. unphysical regions where the potential
energy lies below the (physically sound) global potential minimum.  As such regions pose
problems both for the PES fitting (where they introduce artificial correlations between
the DOFs) and for the quantum calculations (where they may act as a trap for the
wavefunction), it is advisable to repair or avoid those regions. While Yu \cite{yu_rigorous_2006}
replaced all negative energies with a large positive value\footnote{%
We note that this procedure might shift the GS energy upwards.}
so that the wavefunction avoids
the unphysical regions, Peláez et al. \cite{pel13:014108, pel14:42}
carefully reduced the coordinate ranges so that no regions of negative energy can be encountered.
However, unphysically distorted PES regions could not be excluded to be present in the grid,
which might lead to unphysical correlation between the DOFs.
Here we follow Peláez' approach, and reuse the coordinate ranges as well as other
parameters defining the primitive grid from the previous works
\cite{pel14:42,pel17:100}.
These parameters are listed in Table \ref{tab:dvr}.

\begin{table}
\centering
\begin{tabular}{l|lrl}
DOF & DVR & $N$ & Range \\
\hline
$r_1$ & HO & 13 & $[1.4, 2.4]$ \\
$r_2$ & HO & 13 & $[1.4, 2.4]$ \\
$R$ & HO & 11 & $[4.15, 5.5]$ \\
$x$ & HO & 10 & $[-0.8, 0.8]$ \\
$y$ & HO & 10 & $[-0.8, 0.8]$ \\
$z_\text{red}$ & HO & 20 & $[-0.5, 0.5]$ \\
$u_1$ & sin & 12 & $[-0.8, 0.35]$ \\
$u_2$ & sin & 12 & $[-0.35, 0.8]$ \\
$\phi$ & exp & 21 & $[0, 2\pi)$ \\
\end{tabular}
\caption{%
Definition of the primitive grid. The DVR column indicates the type of the
discrete variable representation used for each DOF: harmonic oscillator (HO),
sine (sin), or exponential (exp) DVR. $N$ is the number of grid points for each DOF.
Instead of the angles $\theta_i$ we use
the variables $u_i = \cos \theta_i$. All distance coordinates ($r_1$, $r_2$, $R$,
$x$, $y$) are measured in bohr, and $\phi$ in radians.
$u_1$, $u_2$, and $z_\text{red} = z/(R-2d_0)$ (see text) are dimensionless.
\label{tab:dvr}}
\end{table}

\section{Results}
\label{sec:results}

Using four different variants of Potfit, we have computed the ground
state (GS) of the \HHHOO{} ion, for which the PES3C potential energy surface
described in Section \ref{sec:h3o2} was used.
Our investigation focuses on how the accuracy of the PES
fit influences the runtime of the GS computation.

\subsection{Computational Setup}

\begin{figure}
\centering
\includegraphics[width=8cm]{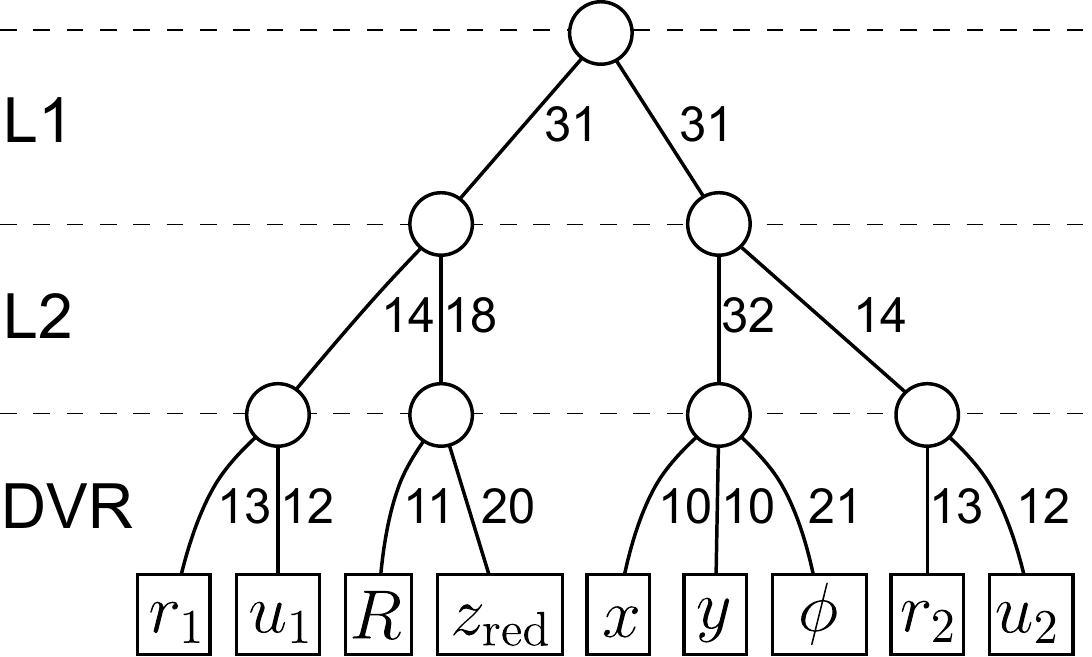}
\caption{%
ML-MCTDH wavefunction tree used for our 4-mode computations.
The edges are labeled with the number of basis functions for each mode, \ie the number of
DVR grid points for the primitive modes (bottom), and the number of
SPFs for layer 2 (L2) and layer 1 (L1) modes.
\label{fig:tree4mode}}
\end{figure}

To discretize the system, we employed the same primitive basis
as in Ref.\ \cite{pel14:42}, and the basis parameters are listed in Table \ref{tab:dvr}.
We also reuse the kinetic energy operator from \cite{pel14:42}
which was derived as an analytic expression using the
TANA software package \cite{ndo12:034107}.
However, we used a different mode combination setup than in \cite{pel14:42}
because here we perform most computations with ML-MCTDH instead of MCTDH,
and we found that the original mode combination scheme led to suboptimal
performance of ML-MCTDH, and that a different scheme yielded an improved
balance of the ML-MCTDH tree.
The primitive combined modes we used for our MCTDH calculations are:
$(r_1, u_1)$, $(R,z_\text{red})$, $(x,y,\phi)$, and $(r_2,u_2)$.
For our ML-MCTDH calculations, the first two and the last two of these primitive modes were
then combined for the upper layer.  The resulting ML-MCTDH tree is depicted in
Fig.\ \ref{fig:tree4mode}.

\begin{figure}
\centering
\includegraphics[width=8cm]{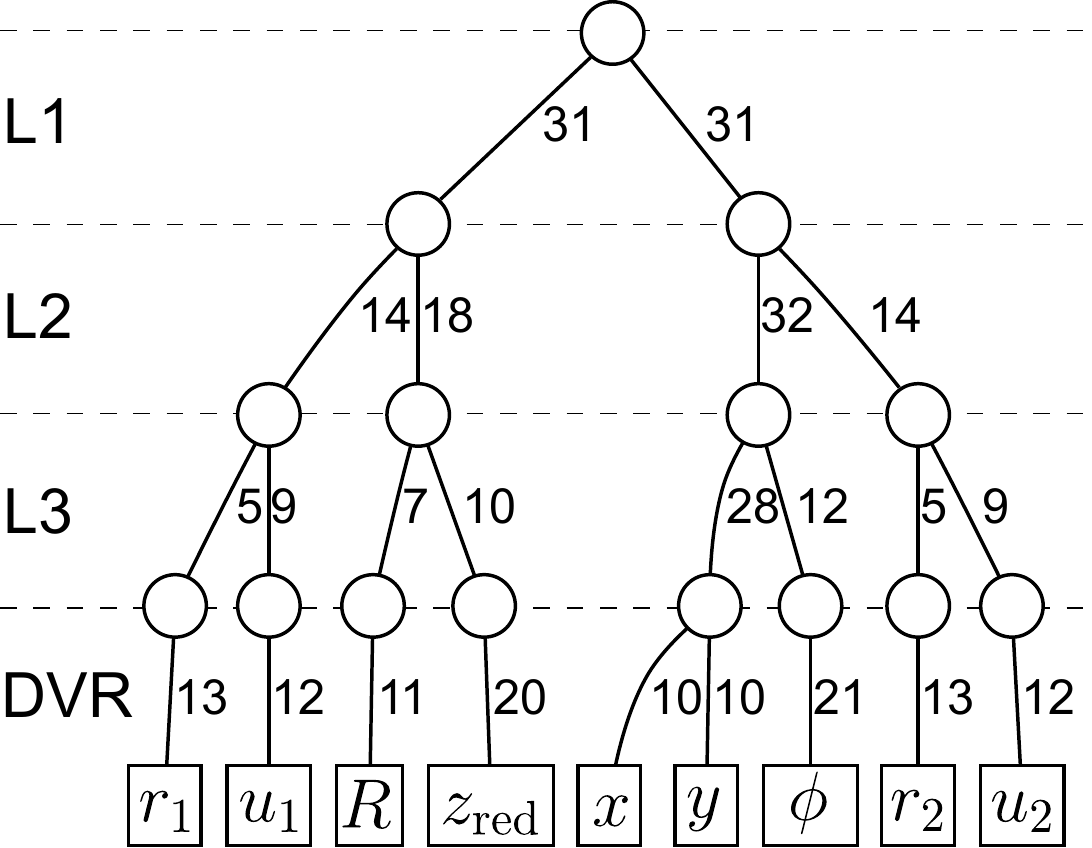}
\caption{%
ML-MCTDH wavefunction tree used for our 8-mode computations.
The edges are labeled as in Fig.\ \ref{fig:tree4mode}, and a third layer
of SPFs (L3) has been added.
\label{fig:tree8mode}}
\end{figure}

In addition to this setup with four modes, we have perfomed a set of
ML-MCTDH computations with eight modes, where all primitive modes
contain only one DOF except for the combined mode $(x,y)$.
The resulting ML-MCTDH tree is shown in Fig.\ \ref{fig:tree8mode}.
We note that, at the desired level of accuracy for the wavefunction,
MCTDH computations with eight modes would be much too costly.
Therefore all 8-mode computations were carried out with ML-MCTDH.

To obtain the system's ground state, we used the improved relaxation method
with MCTDH as well as regular relaxation (\ie propagation in negative imaginary
time) with ML-MCTDH.  All computations started with the same initial wavefunction,
which was defined as a Hartree product of Gaussians around the energy-optimized geometry.
In order to quickly obtain a state relatively close to the ground state, we first performed a
set of short relaxation runs, using gradually increasing number of SPFs and
increasing integrator accuracy.   The state obtained in this initial phase was then
subjected to a long relaxation run (nominal propagation time 1000\ fs), using high integrator accuracy and
the number of SPFs as indicated in Figs.\ \ref{fig:tree4mode} and \ref{fig:tree8mode}.
We note that with these settings, the lowest natural populations for all modes never
exceeded $10^{-7}$, which shows that our wavefunction representation is highly accurate.
Finally, to check that the GS energy is converged, the number of SPFs for all modes was
slightly increased, and a further short relaxation run was performed. In all our computations,
the energy dropped at most by $0.01\,\ic$, which indicates that convergence had been achieved.
All relaxation runtimes reported below refer to the long main relaxation run over 1000\ fs,
as this was the part which dominated the runtime of the ground state computation.

\begin{figure}
\centering
\includegraphics[width=8cm]{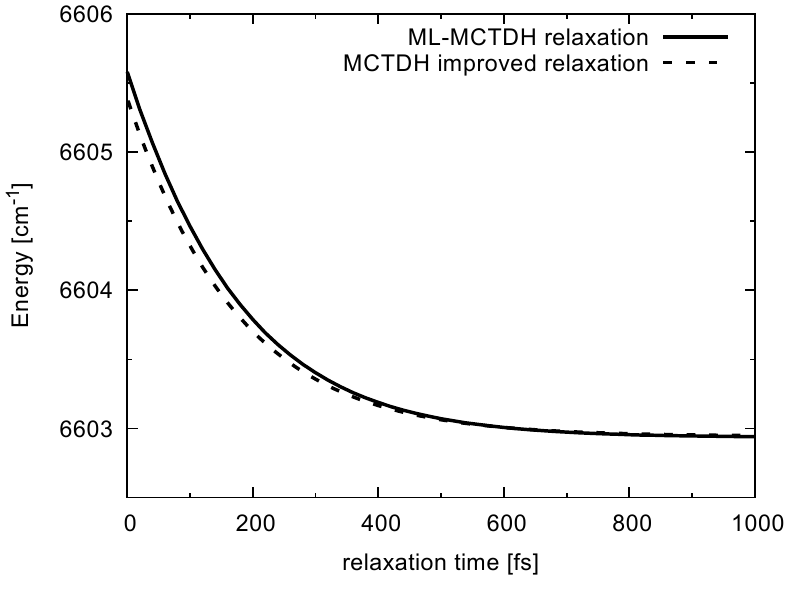}
\caption{%
Evolution of the energy expectation value $\langle \Psi | \hat{H} | \Psi \rangle$
over the course of the relaxation, for one exemplary MCTDH improved relaxation calculation
and one exemplary ML-MCTDH relaxation calculation.
\label{fig:E-vs-t}}
\end{figure}

For MCTDH, we used the same number of SPFs as for L2 of the ML-MCTDH tree,
\ie 14/18/32/14, resulting in an $A$-vector of 112,896 elements.
As discussed in Section \ref{sec:mctdh}, the improved relaxation method only uses
propagation for the SPFs but diagonalization to obtain the $A$-vector. Nevertheless, we found
that the convergence behaviour towards the GS differs very little between improved
relaxation runs and regular relaxation runs, as demonstrated by the relaxation
profiles shown in Fig.\ \ref{fig:E-vs-t}.  Note that both methods indeed need
about 1000\ fs to converge to the GS.  This allows us to make a fair comparison between
the runtimes for these two relaxation calculations, despite their algorithmic differences.

\subsection{PES fits}

We have produced a series of fits of the PES3C surface, using
four different variants of Potfit:
\begin{enumerate}
\item
\textbf{MGPF.}
We used multi-grid Potfit (MGPF) in its top-down variant, in which
a very large coarse grid is chosen, essentially utilizing every second fine grid point.
This produced a fit of very high accuracy
with 49, 338, and 49 SPPs for the modes $(r_1,u_1)$, $(x,y,\phi)$, and $(r_2,u_2)$,
respectively. As in \cite{pel14:42}, the mode $(R,z_\text{red})$ was contracted.
This fit itself would be too large to perform MCTDH calculations, but it can be
used as a starting point for creating more reasonably sized fits by truncating
the number of SPPs for each mode, based on a threshold for their natural weights.
We thus produced three fits at different levels of accuracy, which we refer to by their
relative quality: low (using 14/22/14 SPPs), medium (18/38/18), and high (23/56/23).
The resulting fits in SOP form are suitable for both MCTDH and ML-MCTDH calculations.
\item
\textbf{MGPF+MLPF.}
Each of the three MGPF fits was then post-processed by MLPF, namely the first
step of FG-MLPF can be replaced by reading in the SOP fit produced by MGPF,
computing the SPPs for the contracted mode, and then the 
remaining steps of FG-MLPF are completed.  The FG-MLPF implementation can
automatically determine the expansion orders for all modes, based on a
user-specified target accuracy, which is specified as the global RMS error of the fit.
This feature makes use of the error estimate from the neglected eigenvalues of the
potential density matrix, \ie a multi-layer analogon of \refeq{eq:potfit-error}
(see \cite{ott14:014106} for details).  When postprocessing an existing SOP fit,
this target accuracy refers to the \emph{additional} approximation introduced
by MLPF.  We used two options for this post-processing accuracy, $10\ \ic$ and $2\ \ic$.
\item
\textbf{FG-MLPF.}
Unfortunately we were not able to perform FG-MLPF with the
original grid (cf.\ Table \ref{tab:dvr}), which contains $11.2 \cdot 10^9$ points.
Storing the PES on this grid with double-precision floating point numbers requires 84\ GiB of memory,
and the FG-MLPF implementation needs additional storage of similar size.
To make the FG-MLPF fits feasible, we reduced the number of grid points for
the $\phi$ DOF from 21 to 15. This reduces the storage requirements for the PES
to 60\ GiB, making the computation possible on a single machine with 128\ GiB of RAM.
Again, we can choose the desired global RMS error of the FG-MLPF fit,
for which we selected the three target accuracies $10\ \ic$, $2\ \ic$, and $0.1\ \ic$.  The
latter option serves as a reference point, as it reproduces the full-grid PES almost
exactly.
\item
\textbf{RS-MLPF.}
As described in Section \ref{sec:rs-mlpf}, in addition to the target accuracy,
we need to specify the oversampling parameter $q$ for the algorithm.
While the target accuracy controls the expansion orders as in FG-MLPF,
the random sampling incurs an additional approximation error. This error
can be reduced by increasing $q$, which in turn increases the runtime for
the RS-MLPF algorithm.  However, the size of the resulting fit depends
strongly only on the target accuracy but not on $q$, and it is this size
which is essential for the performance of the subsequent ML-MCTDH calculations.
Here we explored the effects of setting $q$ to either 2, 3, 4, or 6.
\end{enumerate}

To judge the actual quality of these fits, we estimated their root-mean-square
(RMS) error with respect to the original potential with a variety of sampling
methods (details are given in \ref{app:errors}):
uniform random sampling, Markov chain Monte Carlo, and classical
molecular dynamics, where the latter two methods take an energy (or temperature)
parameter for defining the Boltzmann distribution to sample over. We found that they
yield very similar results at comparable temperature settings.
Here we list the RMS errors obtained by molecular dynamics
sampling at $T=600\,\text{K}$ (\ie $\kB T = 417\,\ic$), by Monte Carlo
sampling at $\kB T=10,000\,\ic$, and by uniform random sampling (\ie
formally at $T=\infty$), for all the fits that we produced.
See Tables \ref{tab:fits4mode} and \ref{tab:fits8mode} for the
4-mode and the 8-mode fits, respectively.
Each sample consisted of one million points; for uniform random sampling
(which covers the largest region to sample over) we also performed tests
with ten million sampling points, but the resulting RMS errors only differ
by at most 3\%, hence we conclude our sample size to be sufficient. 

We observe the following noteworthy results about the fit quality:
\begin{itemize}
\item
When comparing the original MGPF fits to those postprocessed by MLPF,
we find that setting the additional fit accuracy to $2\,\ic$ leads to no
significant change in the fit errors, even for the high-quality MGPF fit.
Practically, these postprocessed fits have the same quality
level as the original MGPF fits.  Even when using a less accurate setting
of $10\,\ic$ for the additional fit error, the quality of the fit is barely
affected, viz.\ the estimated fit error is at most $1.1\,\ic$ higher than
for the original MGPF fit.
\item
For FG-MLPF, the estimated global RMS error ($T=\infty$) strictly
observes the user-supplied accuracy parameter.  This is by design, as the
algorithm automatically determines the expansion order for each mode
such that the the user-supplied RMS threshold is not exceeded.
Although this error control scheme operates using the \emph{global}
RMS error, we note that the error estimates at lower temperatures
consistently lie well below this global error measure, viz.\ the fit errors
for $\kB T = 417\,\ic$ and $10,000\,\ic$ lie around 30\% and 60\% of the
global error, respectively.
\item
For the 4-mode RS-MLPF fits (Table \ref{tab:fits4mode}), the global RMS error generally exceeds the prescribed
accuracy parameter. This is expected, as the current error control scheme only
takes into account the error caused by truncating the number of SPPs based
on their natural weights, but the additional error caused by computing the
optimal SPPs only approximately is not controlled for.
We note that this additional error (the ``RS error'') can be lowered significantly
by increasing the oversampling parameter $q$.  We find that at $q=6$, the
actual global RMS error lies reasonably close to the prescribed accuracy parameter
(within a factor of two).  Moreover, for the lower temperature
settings the fit errors behave significantly better than the global error, as they appear to be
much less affected by the RS error.  Especially for the setting $\kB T = 417\,\ic$, the fit
error is barely worse than for FG-MLPF at the same target accuracy, even at $q=2$.
For $\kB T = 10,000\,\ic$, the fit error can be brought close to the FG-MLPF result by
choosing $q=3$ or $q=4$.
\item
For the 8-mode RS-MLPF fits (Table \ref{tab:fits8mode}), we found that the fit errors
were much worse than for the 4-mode RS-MLPF fits at the same settings for target
accuracy and $q$.  Moreover, these fits finish much faster than the corresponding 4-mode fits --
up to 40 times faster at low accuracy, but this advantage diminishes for higher accuracy.
Both phenomena can be explained by the fact that
for the 8-mode fits, most of the primitive modes
are very small as they only contain a single DOF, so that even with a large $q$ the number
of PES evaluations for computing their potential density matrices is small (as low as a few hundred). 
This leads to a rather large RS error, which we here aim to reduce by
introducing an additional parameter
$f_\text{min}$ which signifies the minimum number of PES evaluations to be used when computing
the potential density matrix for each mode (technically, this is done by locally adjusting $q$ upwards).
This significantly increases the fit accuracy, often by a factor of two even for $f_\text{min}=10^5$.
For $q=6$ and $f_\text{min}=10^6$ we generally reach the same level of accuracy as the 4-mode fits.
The additional computation time for the fit is relatively modest, viz.\ several seconds (for $f_\text{min}=10^5$)
to a few minutes (for $f_\text{min}=10^6$).
\item
In order to judge the reliability of the RS-MLPF method, we have repeated some of the 8-mode fits
100 times with different random seeds.  The resulting average fit errors and their uncertainties
(measured as the standard deviation) are recorded in Table \ref{tab:fits8mode}.  We note that with $f_\text{min}=0$,
the relative uncertainties are rather large, often exceeding 20\%.  By setting $f_\text{min} \geq 10^5$, the relative
uncertainties can generally be brought down to less than 10\%, especially for the fit errors at lower
temperatures.  This means that setting $f_\text{min}$ not only increases the accuracy of the fit, but also
increases the reliability of the fit result, because it reduces the statistical fluctuations inherent in our
randomized algorithm.
\item
The computational resource requirements for RS-MLPF are relatively low. We stress that the
RS-MLPF timings reported in Tables \ref{tab:fits4mode} and \ref{tab:fits8mode} were obtained
using a single core on an Intel i5-7200U laptop CPU.  Runtimes range from a few seconds to a
few minutes, depending on the desired accuracy, and even the reference fits (target $0.1\,\ic$)
could be completed within one hour.  More importantly, the memory required for running the
RS-MLPF algorithm is usually less than 1\ GiB -- this is a major improvement over the FG-MLPF
algorithm, which required about 120\ GiB of RAM to run.
\item
The size of the RS-MLPF fits (judged by the required number of SPPs) is in general very close
to the size of the corresponding FG-MLPF fit, except that for the top layer (L1), slightly more
SPPs seem to be required.  This phenomenon diminishes with increasing $q$.  Hence a larger
$q$ is not only beneficial for reducing the RS error, but also makes the fit a bit more compact,
which has runtime advantages for ML-MCTDH.
\end{itemize}

\begin{sidewaystable}
\renewcommand{\arraystretch}{0.8}
\centering
\begin{tabular}{lrr|cc|rrr|r@{}l}
\hline
Method & Target & $q$ & \multicolumn{2}{c|}{SPP} & \multicolumn{3}{c|}{$\kB T [\ic]$} &  $t_\text{fit}$[s] \\
\cline{4-5} \cline{6-8}
& $[\ic]$ && L2 & L1 & $417$ & $10,000$ & $\infty$ \\
\hline
\hline
MGPF & && 14,c,22,14 && $8.8$ & $20.0$ & $60.2$ \\
MGPF & && 18,c,38,18 && $3.7$ & $7.9$ & $55.0$ \\
MGPF & && 23,c,56,23 && $1.7$ & $3.9$ & $54.2$ \\ 
\hline
MGPF+MLPF & 10.0 && 14,18,22,14 & 55,55 & $9.3$ & $20.3$ & $60.6$ \\
MGPF+MLPF & 2.0 && 14,32,22,14 & 92,92 & $8.8$ & $19.8$ & $60.2$ \\
MGPF+MLPF & 10.0 && 18,19,38,18 & 60,60 & $4.3$ & $9.0$ & $55.0$ \\
MGPF+MLPF & 2.0 && 18,39,38,18 & 110,110 & $3.7$ & $7.9$ & $55.0$ \\
MGPF+MLPF & 10.0 && 23,20,56,23 & 62,62 & $2.5$ & $5.6$ & $54.3$ \\
MGPF+MLPF & 2.0 && 23,43,56,23 & 121,121 & $1.7$ & $4.0$ & $54.3$ \\
\hline
FG-MLPF & 40.0 && 13,12,24,13 & 31,31 & $12.0$ & $ 23.7$ & $39.3$ & 9573&$^a$ \\
FG-MLPF & 10.0 && 18,17,38,18 & 53,53 & $2.8$ & $5.7$ & $9.7$ & 9924&$^a$ \\
FG-MLPF & 2.0 && 23,22,58,24 & 84,84 & $0.7$ & $1.2$ & $2.0$ & 10364&$^a$ \\
FG-MLPF & 0.1 && 34,33,109,35 & 165,165 & $0.03$ & $0.06$ & $0.10$ & 11446&$^a$ \\
\hline
RS-MLPF & 40.0 & 2 & 12,12,24,13 & 36,36 & $14.8$ & $43.1$ & $101.8$ & 83 \\
RS-MLPF & 40.0 & 3 & 13,12,24,13 & 33,33 & $13.4$ & $30.0$ & $57.7$ & 120 \\
RS-MLPF & 40.0 & 4 & 13,12,24,13 & 32,32 & $13.0$ & $26.9$ & $55.4$ & 153 \\
RS-MLPF & 40.0 & 6 & 13,12,24,13 & 32,32 & $13.0$ & $25.6$ & $46.7$ & 229 \\
RS-MLPF & 10.0 & 2 & 17,17,38,18 & 58,58 & $3.3$ & $11.8$ & $34.4$ & 95 \\
RS-MLPF & 10.0 & 3 & 17,17,38,18 & 55,55 & $3.1$ & $7.9$ & $19.5$ & 151 \\
RS-MLPF & 10.0 & 4 & 17,17,38,18 & 55,55 & $2.9$ & $7.0$ & $16.1$ & 201 \\
RS-MLPF & 10.0 & 6 & 18,17,38,18 & 53,53 & $2.7$ & $6.3$ & $13.1$ & 319 \\
RS-MLPF & 2.0 & 2 & 23,21,57,23 & 94,94 & $0.7$ & $3.0$ & $11.3$ & 115 \\
RS-MLPF & 2.0 & 3 & 22,22,57,23 & 89,89 & $0.7$ & $2.3$ & $6.5$ & 198 \\
RS-MLPF & 2.0 & 4 & 23,22,57,24 & 86,86 & $0.7$ & $1.5$ & $3.7$ & 277 \\
RS-MLPF & 2.0 & 6 & 23,22,57,24 & 87,87 & $0.7$ & $1.3$ & $3.3$ & 477 \\
RS-MLPF & 0.1 & 6 & 34,33,107,34 & 173,173 & $0.03$ & $0.07$ & $0.19$ & 1594 \\ 
\hline
\end{tabular}
\caption{%
Details of our 4-mode fits. The ``target'' column lists the accuracy parameter
used for the MLPF error control. For RS-MLPF, $q$ signifies the oversampling parameter.
The number of SPPs for layers 2 and 1 are listed in the L2 and L1 columns, respectively
(cf.\ Fig.\ \ref{fig:tree4mode}).  The $\kB T$ columns list the estimated fit error at different
temperature settings as described in the text. $t_\text{fit}$ lists the time that it took
to produce the fit.
$^a$ Time for fit only. Calculation of the full PES took 5183s on 16 CPU cores, but had to
be performed only once for all FG-MLPF fits.
\label{tab:fits4mode}}
\end{sidewaystable}

\begin{sidewaystable}
\renewcommand{\arraystretch}{0.8}
\centering
\begin{tabular}{lrrr|ccc|r@{}lr@{}lr@{}l|rr}
\hline
Method & Target & $q$ & $f_\text{min}$ & \multicolumn{3}{c|}{SPP} & \multicolumn{6}{c|}{$\kB T [\ic]$} & $f/10^6$ & $t_\text{fit}$[s]\\
\cline{5-7} \cline{8-13}
& $[\ic]$ & & & L3 & L2 & L1 & $417$ && $10,000$ && $\infty$& & & \\
\hline
\hline
FG-MLPF & $10.0$ &&& 6,6,5,7,25,13,6,6 & 18,17,39,18 & 55,55 & 2.8 && 5.6 && 9.8 && 8030.9 & 8889 \\
FG-MLPF & $2.0$ &&& 6,7,5,8,33,15,6,7 & 24,22,60,24 & 86,86 & 0.7 && 1.2 && 2.0 && 8030.9 & 9044 \\
FG-MLPF & $0.1$ &&& 7,8,7,10,52,15,7,8 & 35,34,112,36 & 168,168 & 0.0&3 & 0.0&6 & 0.1&0 & 8030.9 & 9526 \\
\hline
RS-MLPF & 40.0 & 2 & 0 & 4,5,4,5,16,8,4,5 & 13,11,24,12 & 41,41 & 33.6 & (6.4) & 114.4 & (20) & 340.4 & (117) & 0.3 & 2 \\
RS-MLPF & 40.0 & 2 & $10^5$ & 4,5,4,6,17,10,4,5 & 14,13,24,14 & 42,42 & 18.2 & (1.8) & 58.6 & (5.6) & 186.2 & (37) & 1.5 & 11 \\
RS-MLPF & 40.0 & 3 & 0 & 4,5,4,6,16,8,4,5 & 12,12,24,12 & 38,38 & 23.0 & (2.7) & 60.9 & (6.8) & 151.9 & (22) & 0.6 & 5 \\
RS-MLPF & 40.0 & 3 & $10^5$ & 4,5,4,6,17,10,4,5 & 13,13,25,13 & 39,39 & 17.1 & (3.4) & 38.4 & (5.7) & 93.6 & (11) & 1.9 & 16 \\
RS-MLPF & 40.0 & 6 & 0 & 4,5,4,6,17,8,4,5 & 12,12,24,13 & 35,35 & 18.4 & (1.6) & 38.8 & (3.3) & 83.3 & (9.4) & 2.4 & 20 \\
RS-MLPF & 40.0 & 6 & $10^5$ & 4,5,4,6,17,10,4,5 & 13,12,25,13 & 37,37 & 15.5 & (1.2) & 28.3 & (1.5) & 57.0 & (1.9) & 4.0 & 36 \\
RS-MLPF & $10.0$ & 2 & $0$ & 5,5,5,7,23,9,5,5 & 17,18,38,17 & 64,64  & 8.5 && 35.6 && 108.2 && 1.0 & 10 \\
RS-MLPF & $10.0$ & 2 & $10^5$ & 6,6,5,7,23,13,6,6 & 19,18,39,20 & 64,64 & 4.4 && 19.1 && 70.8 && 2.5 & 23 \\
RS-MLPF & $10.0$ & 3 & $0$ & 5,6,5,7,22,9,5,6 & 17,16,37,17 & 58,58 & 6.2&(0.9) & 21.1&(3.7) & 60.9&(14) & 2.0 & 18 \\
RS-MLPF & $10.0$ & 3 & $10^5$ & 6,6,5,7,24,13,6,6 & 18,17,38,18 & 59,59 & 3.6&(0.2) & 10.6&(0.8) & 31.3&(4.7) & 3.9 & 36 \\
RS-MLPF & $10.0$ & 4 & $0$ & 6,6,4,7,23,10,6,6 & 17,16,37,18 & 58,58 & 4.7 && 14.1 && 38.0 && 3.8 & 35 \\
RS-MLPF & $10.0$ & 4 & $10^5$ & 6,6,5,7,23,13,6,6 & 18,17,37,18 & 57,57 & 3.4 && 8.2 && 18.6 && 5.8 & 52 \\
RS-MLPF & $10.0$ & 6 & $0$ & 6,6,5,7,23,11,5,6 & 17,17,38,18 & 59,59 & 4.1 && 10.6 && 27.0 && 9.6 & 87 \\
RS-MLPF & $10.0$ & 6 & $10^5$ & 6,6,5,7,23,13,6,6 & 18,17,37,18 & 55,55 & 3.3&(0.3) & 7.5&(0.4) & 16.5&(0.9) & 11.7 & 105 \\
RS-MLPF & $10.0$ & 6 & $10^6$ & 6,6,5,7,24,13,6,6 & 18,17,39,18 & 56,56 & 2.7 && 6.6 && 13.2 && 22.2 & 196 \\
RS-MLPF & $2.0$ & 2 & $0$ & 6,6,5,7,29,11,6,7 & 22,22,57,23 & 100,100 & 2.6 && 15.4 && 64.2 && 3.0 & 29 \\
RS-MLPF & $2.0$ & 2 & $10^5$ & 6,7,5,8,33,15,6,7 & 25,24,61,25 & 106,106 & 2.0 && 13.2 && 52.5 && 5.7 & 61 \\
RS-MLPF & $2.0$ & 3 & $0$ & 6,7,5,8,31,11,6,7 & 22,22,58,23 & 94,94 & 1.5 && 8.8 && 40.9 && 6.9 & 63 \\
RS-MLPF & $2.0$ & 3 & $10^5$ & 6,7,5,8,32,17,6,7 & 24,23,59,24 & 92,92 & 0.9 && 3.7 && 17.1 && 10.8 & 99 \\
RS-MLPF & $2.0$ & 4 & $0$ & 6,7,5,8,31,13,6,7 & 23,22,58,23 & 91,91 & 1.1 && 5.2 && 13.4 && 13.5 & 129 \\
RS-MLPF & $2.0$ & 4 & $10^5$ &  6,7,5,8,33,17,6,7 & 24,23,59,24 & 91,91 & 0.8 && 2.5 && 6.7 && 18.6 & 169 \\
RS-MLPF & $2.0$ & 6 & $0$ & 6,7,5,7,30,13,6,7 & 23,21,58,23 & 93,93 & 0.9 && 3.3 && 7.4 && 28.9 & 282 \\
RS-MLPF & $2.0$ & 6 & $10^5$ &  6,7,5,8,32,16,6,7 & 23,22,59,24 & 91,91 & 0.8&(.05) & 1.9&(.15) & 5.0&(0.9) & 36.3 & 328 \\
RS-MLPF & $2.0$ & 6 & $10^6$ &  6,7,5,8,34,17,6,7 & 24,23,58,24 & 91,91 & 0.7 && 1.6 && 3.5 && 50.7 & 455 \\
RS-MLPF & $0.1$ & 6 & $10^6$ &  7,8,7,10,51,21,7,8 & 35,34,110,35 & 174,174 & 0.03 && 0.09 && 0.24 && 217.2 & 2011 \\
RS-MLPF & $0.1$ & 10 & $10^6$ & 7,8,7,10,52,21,7,8 & 34,33,111,35 & 178,178 & 0.03 && 0.07 && 0.23 && 566.1 & 5114 \\
\hline
\end{tabular}
\caption{%
Details of our 8-mode fits. Columns are as in Table \ref{tab:fits4mode}, with an additional column L3
for the number of SPPs for layer 3 (cf.\ Fig.\ \ref{fig:tree8mode}).  For RS-MLPF, $f_\text{min}$ lists
the minimum number of PES evaluations per mode. $f$ is the actual total number of PES evaluations
needed to produce the fit.  Numbers in parentheses signify absolute uncertainties (one standard deviation) obtained
from performing the same fit 100 times with different random seeds.
\label{tab:fits8mode}}
\end{sidewaystable}

\subsection{Ground State Relaxation}

For all the PES fits discussed in the previous section,
we have computed the ground state (\ie zero-point) energy $E_\text{GS}$
of the \HHHOO{} system.
While we are interested in how the runtime for the ground state computation
(performed with MCTDH or ML-MCTDH) depends on the type and the 
accuracy of the PES fit, we note that this runtime mostly depends on the
size of the PES fit, \ie on the number of SPPs used for each mode.
Many of our PES fits, especially the RS-MLPF fits, exhibit very similar size
in this regard, therefore we didn't find it necessary to perform a full
relaxation run for every single one of our PES fits.  Faced with limited
computational resources, we instead opted
to perform relaxation runs only for the most accurate RS-MLPF fits,
and use the obtained ground state wavefunction to compute $E_\text{GS}$
for the other, less accurate fits with a perturbation theory approach.
Namely, let $V$ be the PES fit used for computing the ground state
$\Psi_\text{GS}$, and let $\tilde{V}$ be another PES fit with its accompanying
ground state $\tilde{\Psi}_\text{GS}$.  Then we can estimate
\begin{equation}
\tilde{E}_\text{GS} = \langle \tilde{\Psi}_\text{GS} | \hat{T} + \tilde{V} | \tilde{\Psi}_\text{GS} \rangle
\approx \langle \Psi_\text{GS} | \hat{T} + \tilde{V} | \Psi_\text{GS} \rangle
\end{equation}
where $\hat{T}$ denotes the kinetic energy operator.  We have verified that this
is a very accurate estimate by computing the GS energy for the same PES fit with
all GS wavefunctions of the other fits, and found that the energies differed at
most by $0.02\,\ic$.
The obtained GS energies and, where applicable, relaxation runtimes are
listed in Tables \ref{tab:relax4mode} and \ref{tab:relax8mode} for our
4-mode and 8-mode computations, respectively.
All relaxation calculations where performed on a compute node with
two Intel Xeon E5-2670 CPUs, using shared memory parallelization
with 16 CPU cores.

\begin{table}
\renewcommand{\arraystretch}{0.8}
\centering
\begin{tabular}{lrrlrr@{}l}
\hline
Fit & Target$[\ic]$ & $q$ & Method$^a$ & $E_\text{GS}[\ic]$ & $t_\text{rlx}[h]$ \\
\hline
\hline
\multicolumn{6}{c}{MGPF low (SPP = 14,c,22,14)} \\
original &&& imprlx & 6600.42 & 1.7 \\
original &&& rlx,ML & 6600.42 & 12.9 \\
+MLPF & 10.0 && rlx,ML & 6600.61 & 2.8 \\
+MLPF & 2.0  && rlx,ML & 6600.49 & 3.0 \\
\hline
\multicolumn{6}{c}{MGPF medium (SPP = 18,c,38,18)} \\
original &&& imprlx & 6602.88 & 3.6 \\
original &&& rlx,ML & --- & 32\phantom{.0}&$^b$ \\
\hline
\multicolumn{6}{c}{MGPF high (SPP = 23,c,56,23)} \\
original &&& imprlx & 6602.95 & 7.6 \\
original &&& rlx,ML & --- & 68\phantom{.0}&$^b$ \\
+MLPF & 10.0 && rlx,ML & 6602.75 & 3.7 \\
+MLPF & 2.0 && rlx,ML & 6602.94 & 4.0 \\
\hline
FG-MLPF & 40.0 && PT,ML & 6600.39 & --- \\
FG-MLPF & 10.0 && rlx,ML & 6602.80 & 2.4 \\
FG-MLPF & 2.0 && rlx,ML & 6603.17 & 3.7 \\
FG-MLPF & 0.1 && rlx,ML & 6603.29 & 7.0 \\
\hline
RS-MLPF & 40.0 & 2 & PT,ML & 6604.82& --- \\
RS-MLPF & 40.0 & 3 & PT,ML & 6599.86& --- \\
RS-MLPF & 40.0 & 4 & PT,ML & 6599.05& --- \\
RS-MLPF & 40.0 & 6 & rlx,ML & 6597.86 & 2.0 \\
RS-MLPF & 10.0 & 2 & PT,ML & 6603.28& --- \\
RS-MLPF & 10.0 & 3 & PT,ML & 6602.79& --- \\
RS-MLPF & 10.0 & 4 & PT,ML & 6602.64& --- \\
RS-MLPF & 10.0 & 6 & rlx,ML & 6602.82 & 2.8 \\
RS-MLPF & 2.0 & 2 & PT,ML & 6603.14& --- \\
RS-MLPF & 2.0 & 3 & PT,ML & 6603.32& --- \\
RS-MLPF & 2.0 & 4 & PT,ML & 6603.23& --- \\
RS-MLPF & 2.0 & 6 & rlx,ML & 6603.18 & 3.7 \\
RS-MLPF & 0.1 & 6 & rlx,ML & 6603.29 & 6.7 \\
\hline
\end{tabular}
\caption{%
Ground state energy $E_\text{GS}$ and runtimes $t_\text{rlx}$ for the relaxation computations,
using our 4-mode PES fits as defined by fit method, target accuracy, and oversampling parameter $q$.
$^a$ Method for obtaining $E_\text{GS}$: imprlx = by improved relaxation, rlx = by regular relaxation,
PT = by perturbation theory (see text), ML = using ML-MCTDH.
$^b$ Estimated (see text).
\label{tab:relax4mode}}
\end{table}

\begin{table}
\renewcommand{\arraystretch}{0.8}
\centering
\begin{tabular}{lrrrlr@{}lr}
\hline
Fit & Target$[\ic]$ & $q$ & $f_\text{min}$ & Method & $E_\text{GS}[\ic]$ && $t_\text{rlx}[h]$ \\
\hline
\hline
FG-MLPF & 10.0 &&& rlx,ML & 6602.45 && 2.3 \\
FG-MLPF & 2.0 &&& rlx,ML & 6603.10 && 3.4 \\
FG-MLPF & 0.1 &&& rlx,ML & 6603.29 && 7.5 \\
\hline
RS-MLPF & 40.0 & 2 & 0 & PT,ML & 6600.32 & (6.16) & ---\\
RS-MLPF & 40.0 & 2 & $10^5$ & PT,ML & 6600.27 & (2.76) &---\\
RS-MLPF & 40.0 & 3 & 0 & PT,ML & 6601.79 & (4.38) &---\\
RS-MLPF & 40.0 & 3 & $10^5$ & PT,ML & 6598.96 & (2.95) &---\\
RS-MLPF & 40.0 & 6 & 0 & PT,ML & 6599.94 & (2.85) &---\\
RS-MLPF & 40.0 & 6 & $10^5$ & rlx/PT,ML & 6598.78 & (1.55) & 1.4 \\
RS-MLPF & 10.0 & 2 & 0 & PT,ML & 6604.46 && --- \\
RS-MLPF & 10.0 & 2 & $10^5$ & PT,ML & 6603.11 && --- \\
RS-MLPF & 10.0 & 3 & 0 & PT,ML & 6602.78&(1.12) & --- \\
RS-MLPF & 10.0 & 3 & $10^5$ & PT,ML & 6602.90&(0.46) & --- \\
RS-MLPF & 10.0 & 4 & 0 & PT,ML & 6602.41 && --- \\
RS-MLPF & 10.0 & 4 & $10^5$ & PT,ML & 6603.02 && --- \\
RS-MLPF & 10.0 & 6 & 0 & PT,ML & 6603.70 && --- \\
RS-MLPF & 10.0 & 6 & $10^5$ & PT,ML & 6602.43&(0.40) & --- \\
RS-MLPF & 10.0 & 6 & $10^6$ & rlx,ML & 6602.34 && 2.3 \\
RS-MLPF & 2.0 & 2 & 0 & PT,ML & 6603.02 && --- \\
RS-MLPF & 2.0 & 2 & $10^5$ & PT,ML & 6603.00 && --- \\
RS-MLPF & 2.0 & 3 & 0 & PT,ML & 6603.35 && --- \\
RS-MLPF & 2.0 & 3 & $10^5$ & PT,ML & 6603.27 && --- \\
RS-MLPF & 2.0 & 4 & 0 & PT,ML & 6602.97 && --- \\
RS-MLPF & 2.0 & 4 & $10^5$ & PT,ML & 6603.25 && --- \\
RS-MLPF & 2.0 & 6 & 0 & PT,ML & 6603.21 && --- \\
RS-MLPF & 2.0 & 6 & $10^5$ & PT,ML & 6603.15&(0.06) & --- \\
RS-MLPF & 2.0 & 6 & $10^6$ & rlx,ML & 6603.13 && 3.6 \\
RS-MLPF & 0.1 & 6 & $10^6$ & PT,ML & 6603.29 && --- \\
RS-MLPF & 0.1 & 10 & $10^6$ & rlx,ML & 6603.29 && 8.2 \\
\hline
\end{tabular}
\caption{%
Ground state energy $E_\text{GS}$ and runtimes $t_\text{rlx}$ for the relaxation computations,
using our 8-mode PES fits as defined by fit method, target accuracy, oversampling parameter $q$,
and minimum number of PES evaluations per mode $f_\text{min}$.
Method is as in Table \ref{tab:relax4mode}.
Numbers in parentheses are uncertainties (one standard deviation) obtained by performing the
same fit 100 times with different random seeds.
\label{tab:relax8mode}}
\end{table}

\begin{figure}
\centering
\includegraphics[width=8cm]{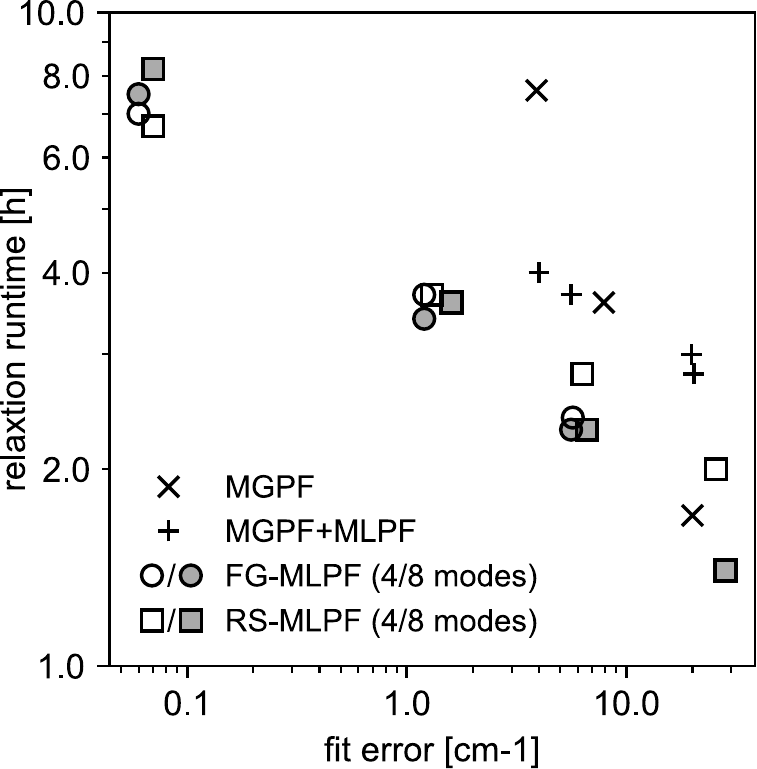}
\caption{%
Runtime for the relaxation calculation vs.\ accuracy of the PES fit.
For MGPF, runtime is for the MCTDH improved relaxation run.
Otherwise, runtime is for the ML-MCTDH regular relaxation run.
The error of the PES fit is estimated by Metropolis-Hastings sampling
over $10^6$ points at an energy of $\kB T = 10,000\,\ic$.
\label{fig:trlx-vs-fiterr}}
\end{figure}

The runtime results for the relaxation calculations are summarized in
Fig.\ \ref{fig:trlx-vs-fiterr}.  We elected to plot the relaxation runtime against
the Monte Carlo estimated fit error at $\kB T = 10,000\,\ic$, as this is the
temperature setting which adequately covers the PES regions where the
GS wavefunction has appreciable population (since $E_\text{GS} \approx 6600\,\ic$).
We note that the runtime for the MCTDH improved relaxation calculations
with the MGPF fits (denoted by the cross symbols) increases rather rapidly
with increasing fit accuracy (\ie with decreasing fit error), whereas the
runtimes for ML-MCTDH relaxations with the MLPF fits only increase moderately so;
namely, using an MLPF fit that is ten times more accurate only roughly doubles
the relaxation runtime.
The reason for this very different scaling behaviour between
the MGPF and MLPF fits lies in the different operator format: 
MGPF produces fits in SOP format, while the MLPF methods produce
fits in MLOp format.  As discussed in Section \ref{sec:pesfit}, the SOP
format necessarily requires a quickly increasing number of terms for
higher accuracy fits.  In \cite{ott14:014106}, it was estimated that the
MLOp format should lead to an asymptotically better scaling behaviour,
at least under moderate assumptions for the expansion orders of
the MLPF fit.  Our new results fully confirm these estimations.

Furthermore, we note that postprocessing the MGPF fit by MLPF
is here detrimental at low accuracy fit settings, roughly doubling
the runtime for the ML-MCTDH relaxation compared to the MCTDH improved
relaxation.  This is mostly due to the fact that improved relaxation is
algorithmically superior to regular relaxation, and due to additional
overhead required by the ML-MCTDH algorithm compared to MCTDH.
However at high accuracy fit settings, postprocessing the MGPF fit
by MLPF becomes benefecial, roughly halving the relaxation runtime.
That is, the better scaling induced by the MLOp format offsets the
inferior algorithmic performance of regular relaxation compared
to improved relaxation.
This effect is even more pronounced when using the direct MLPF
fits (\ie FG- or RS-MLPF), which at all accuracy settings lead to better
runtimes than the MLPF-postprocessed MGPF fits.
Even at low fit accuracy, the direct MLPF fits lead to relaxation runtimes
that can compete with the improved relaxation runtimes for the MGPF fit.
This computational advantage enables us to even perform the full
relaxation run for our reference PES fits (target accuracy $0.1\,\ic$)
in reasonable time ($\sim 8$ hours).  We estimate that an MGPF fit of
similar accuracy would result in a runtime of well over 100 hours, even
with improved relaxation.

We point out that the different scaling behaviour between SOP-format
(MGPF) and MLOp-format (MLPF) fits is indeed due to the different PES
format, and not caused by algorithmic differences between MCTDH improved
relaxation (used with the SOP fits) and ML-MCTDH regular relaxation
(used with the MLOp fits).  To prove this, we have performed additional
ML-MCTDH relaxations with the SOP-format MGPF fits (see
Table \ref{tab:relax4mode}).
Unfortunately, due to the long runtime
of these computations, we were unable to perform a full 1000\ fs relaxation run
except for the low accuracy MGPF, and thus we estimated the full runtime by
extrapolating the runtime from the partial runs.
We observe that these ML-MCTDH relaxation runtimes are
consistently about eight times larger than the corresponding MCTDH improved
relaxation runtimes, regardless of the MGPF accuracy.  That is, the SOP-format
PES shows the same unfavorable scaling behaviour, independent of the
relaxation method.  These calculations also provide a direct measure for
the computational advantage that one can gain by replacing a SOP-format
PES by an MLOp-format PES with similar accuracy: namely, the ML-MCTDH
computation can be sped up between $\sim 4$ and $\sim 20$ times when
postprocessing the MGPF fit with MLPF (at negligible loss of accuracy),
and the gains rapidly increase with the accuracy of the fit.

\begin{figure}
\centering
\includegraphics[width=8cm]{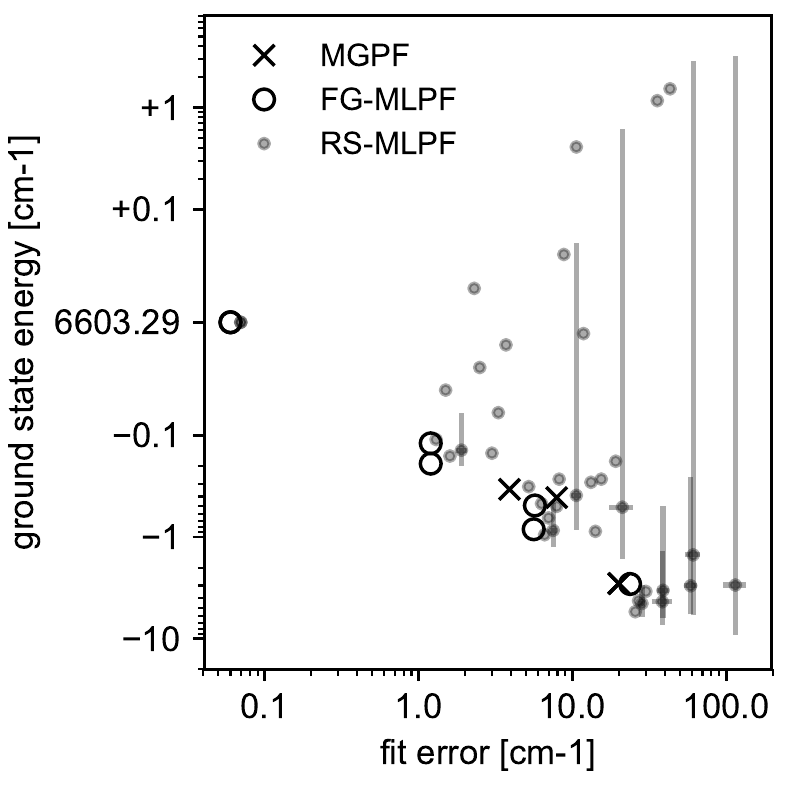}
\caption{%
Ground state energy (obtained by relaxation) vs.\ accuracy of the PES fit.
Data from the RS-MLPF fits are shown by the small grey dots and,
where available, their uncertainties are indicated by error bars (one
standard deviation).
The error of the PES fit is as in Fig.\ \ref{fig:trlx-vs-fiterr}.
Note that the GS energy axis uses linear scaling in the range
$6603.29 \pm 0.10\,\ic$ and logarithmic scaling outside this range.
\label{fig:Egs-vs-fiterr}}
\end{figure}

Finally, we investigate how the GS energy depends on the accuracy
of the PES fit.  Our results are summarized in Fig.\ \ref{fig:Egs-vs-fiterr},
again employing the Monte Carlo error estimate at $\kB T = 10,000\,\ic$.
Our reference PES fits consistently yield a value of $E_\text{GS} = 6603.29\,\ic$.
Fits at lower accuracy deviate from this result; the lower the fit accuracy,
the higher the deviation.  We find that a fit error of $\Delta V$
approximately translates to a GS energy deviation of $\Delta E_\text{GS} = 0.1\, \Delta V$.
Moreover, focusing on the deterministic PES fitting methods (\ie MGPF and FG-MLPF),
we observe a systematic deviation of the GS energy towards lower values.
The GS energies obtained with RS-MLPF also appear to cluster at energies
below the reference result, though this is partly obscured by the fluctuations
induced by the random sampling, which occasionally push $E_\text{GS}$ above
the reference value.  Currently we can not offer a satisfactory explanation for
this apparently systematic deviation.

We believe that our reference result of $E_\text{GS} = 6603.29\,\ic$ is fully
converged to within $0.1\,\ic$ with respect to both the PES fit accuracy
and the wavefunction accuracy.  However, this result holds only for the
primitive basis currently used (cf.\ Table \ref{tab:dvr}).  It is likely that increasing
the coordinate ranges or the number of grid points, or changing the $d_0$
parameter, will change the value of $E_\text{GS}$ by more than $0.1\,\ic$.
Additionally, the PES itself and the electronic structure calculations which
it is based on have errors that are significantly larger than the errors
caused by our more accurate PES fits.  Nevertheless, here we have shown
that the additional fitting error that is necessary for transforming the PES
into a form suitable for ML-MCTDH can be virtually eliminated completely,
at least for systems with 9 DOFs.

\section{Conclusions and Outlook}
\label{sec:conclusion}

We have investigated how the runtime for (ML-)MCTDH calculations
depends on the accuracy of the fit for the potential energy surface (PES),
using relaxation to the ground state of the \HHHOO{} system as
a benchmark.  While it is expected that more accurate PES fits
lead to larger runtimes, we find that how strongly the runtime increases
with the fit accuracy depends on the nature of the fitting method.
We consistently find that PES fits in multi-layer operator (MLOp)
format only lead to a modest increase of the relaxation runtimes,
in contrast to PES fits in sum-of-products (SOP) format which exhibit
a rather rapid increase of the relaxation runtime with the fit accuracy.

As the MLOp format can only be used with ML-MCTDH wavefunctions,
we performed the corresponding ground state computations with
the regular relaxation method (\ie propagation in negative imaginary time).
In contrast, fits in SOP format can be used with the improved relaxation
method available for MCTDH wavefunctions, which is algorithmically superior
to the regular relaxation method.  Despite this algorithmic disadvantage,
we find that the more favorable scaling behaviour of the MLOp fits enables us
to outperform the MCTDH improved relaxation method even at medium
settings for the fit accuracy.  Even with our most accurate PES fits, which were
designed with a targeted global RMS error of around $0.1\,\ic$, the ML-MCTDH
relaxation run converged with modest computational resources (less than ten
hours on 16 CPU cores).  Using these fits, we obtained a zero-point energy
of $6603.3\,\ic$ for the  \HHHOO{} ground state.  This agrees well
with previously reported results of $6605 \pm 5 \ic$ obtained with
the diffusion Monte Carlo method \citep{mccoy_quantum_2005}.

In order to obtain the PES fits in MLOp format, we have designed
a novel variant of the multi-layer Potfit method \cite{ott14:014106}
in which integrations over the full product grid are replaced by
Monte Carlo integrations.  The resulting method,
termed ``random sampling multi-layer Potfit`` (RS-MLPF),
produces PES fits
that are parametrized by a target accuracy (for the global root-mean-square error)
and an oversampling parameter $q$ which controls the number of PES evaluations
used for the Monte Carlo integrations.
Using a prototype implementation of RS-MLPF, we have produced a large number
of PES fits to investigate how its accuracy depends on these parameters, where
the actual fit accuracy was assessed via Monte Carlo as well as molecular dynamics
sampling methods.
Compared to the original ``full-grid'' MLPF,  RS-MLPF requires much less
computational resources (only a single CPU core and less than 1\ GiB of RAM), but
produces larger fit errors due to additional errors from the random sampling.
Hence an RS-MLPF fit will usually miss the prescribed target accuracy, though we
found that setting a large enough $q$ (\eg $q=3$ or $4$)
brings the fit error down to below twice the target accuracy, thus
enabling us to produce PES fits with good \emph{a priori} estimates
for its accuracy.
However, we observed that RS-MLPF occassionally needs some additional tuning
regarding the number of PES evaluations used, and further investigations on
other and larger systems will be needed to develop fully reliable heuristics for
choosing the parameters of the method.

For the 9-dimensional system under study here, the RS-MLPF method allows
us to produce PES fits that are virtually exact, \ie their fit error is tiny
(much below $1\ \ic$) compared to the actual accuracy of the PES. Moreover,
these PES fits can be obtained with modest computational resources,
and they are fully usable with ML-MCTDH, as they don't cause excessive
runtimes for the relaxation or propagation.
These features make us very optimistic that the method will be applicable
to larger systems with up to about 20 DOFs.  While we don't expect to be able to
reproduce the extreme level of accuracy which we obtained for the 9D
system here, fits with good accuracy (\ie below $\sim 10\ \ic$
in the low-energy region) seem to be in reach with RS-MLPF, though
additional work regarding its implementation (especially parallelisation)
will be required.  The combination of RS-MLPF and ML-MCTDH will thus
yield a method for quantum molecular dynamics on general potential
energy surfaces with unprecedented accuracy and performance.

\section*{Acknowledgements}

This research did not receive any specific grant from funding agencies in the public, commercial, or not-for-profit sectors.
Part of this research was carried out while the authors were at the
Theoretical Chemistry Group, Institute for Physical Chemistry,
University of Heidelberg, Germany.  We gratefully acknowledge the
use of their computational facilities for performing the MGPF and
the FG-MLPF fits.
The majority of the relaxation calculations were performed on
computational facilities kindly provided by the Computational
Biophysics group of Prof. Wang Yi at the Chinese University of Hong Kong.
FO thanks Markus Schröder for discussing algorithmic details of
the Monte-Carlo Potfit approach.
Finally, the authors would like to thank Hans-Dieter Meyer for his
continuous friendship, advice and support in scientific as well as in personal matters.

\appendix

\section{Estimating the PES Fitting Error}
\label{app:errors}

The root-mean-square (rms) error of the potential fit $\tilde{V}(q)$
with respect to the original potential $V(q)$ is given by
\begin{equation}
\Delta_{\text{rms}} V \,=\, \left[ \int \text{d}q \, w(q) \left( \tilde{V}(q) - V(q) \right)^2 \right]^{1/2}
\end{equation}
where $w(q)$ is a weight function, which is non-negative and fulfills $\int \text{d}q \, w(q) = 1$.
Replacing the integral by a summation over \emph{all} grid points results in an expression
that is too expensive to evaluate, due to the enormous size of the full product grid.
To estimate the rms error, the integral is instead replaced by  a summation over a
\emph{sample} of grid points, $S = \{ q_i \,|\, 1 \leq i \leq N \}$, where the samples are
drawn according to the probability distribution $w(q)$.

To assess the global accuracy of the fit, we employ uniform random sampling, \ie all grid
points are equally likely to be sampled, corresponding to $w(q) = \text{const}$.
Alternatively, to assess the accuracy of the fit in a specific energy region, we gather the
sample $S$ from a Boltzmann distribution, \ie using
\begin{equation}
w(q) \,=\, \exp( -\beta V(q) ) \left / \int \text{d}q' \, \exp( -\beta V(q') ) \right.
\end{equation}
where $\beta = 1 / \kB T$ depends on a temperature parameter $T$, and
$\kB$ denotes the Boltzmann constant.

The Boltzmann sampling can be performed
using a Markov chain Monte Carlo (MCMC) approach with the Metropolis-Hastings algorithm
\cite{met53:1087,has70:97}.
For the 9D system under study here, this approach works adequately.
However, for larger dimensionality it is known that this algorithm performs
more and more poorly, as the random walkers which are used to explore the
conformation space take more and more time to cover the relevant region
defined by the probability distribution, hence it becomes difficult to reach
satisfactory accuracy.  As an alternative to the MCMC approach,
we here introduce the option of performing the Boltzmann sampling by using
classical molecular dynamics (MD),
which has become the standard approach for sampling complex systems with
higher dimensionality, such as proteins or membranes \cite{Shaw10}.

To perform the MD simulations, we employed the standard MD package NAMD \cite{phillips_scalable_2005}
where we overrode the standard calculation of the atomic forces with a finite difference gradient of the PES3C
potential (since the PES3C routine doesn't offer analytical gradients).  Starting from the potential minimum as
the initial conformation, we simulated an NVT ensemble using the Langevin thermostat \cite{PhysRevA.33.3628}
for one million steps.  Note that the PES3C potential is invariant with respect to permutations of the
hydrogen or the oxygen atoms, and NAMD preserves this invariance as the MD simulation is carried out
in Cartesian coordinates.  However, our PES fit is given in internal coordinates, which do not obey the
permutational invariance.  Therefore, when transforming the conformations generated by NAMD from
Cartesian to internal coordinates, we must make a choice for assigning the three hydrogens according to
the definition of the internal coordinates (see Fig.\ \ref{fig:h3o2}). Our choice was done as follows:
First, for each oxygen, find the hydrogen closest to it. This defines the two OH moieties. Then, the remaining
hydrogen is defined as the bridging H. We found that such an assignment was always possible for all the
conformations generated by the MD simulation. After the transformation to internal coordinates, each
conformation was replaced by the nearest conformation on the product grid, as our PES fits are strictly
only defined on the grid points. It was necessary to drop a small number of conformations which came
to lie outside the coordinate ranges given in Table \ref{tab:dvr}.

As a remark, we note that the thermostat does not fix the system temperature always at
a given value, but rather keeps the average temperature around the given value. That
is, the instantaneous temperature always fluctuates around the desired value.
Due to the small system size (only 5 atoms) this fluctuations can be large, but the
average temperature is well maintained around the prescribed value. For instance,
we performed an MD simulation with a prescribed $T = 300\,\text{K}$ which
yielded an averaged temperature of $\bar{T} = 300.1 \,\text{K}$
with a standard deviation of $\Delta T = 109 \,\text{K}$.
Similar behaviour was found for $T = 600\,\text{K}$ and $T = 1000\,\text{K}$,
with observed temperatures of $598 \pm 219 \,\text{K}$ and $1008 \pm 366 \,\text{K}$, respectively.
Nevertheless, when comparing the MD $T=600\,\text{K}$ error estimates for our PES fits
with the MCMC estimates at a comparable temperature setting ($\kB T = 400\,\ic$, \ie
$T=576\,\text{K}$), we found very good agreement throughout.  Hence we conclude that
the MD error estimate is a viable alternative
to the MCMC approach, and is likely to 
work even in rather large dimensionality.

\section*{References}

\bibliography{mctdh,refs,physchem,math,local}

\end{document}